\documentclass[11pt]{article}

\usepackage[final]{acl}

\usepackage{times}
\usepackage{latexsym}

\usepackage[T1]{fontenc}

\usepackage[utf8]{inputenc}

\usepackage{microtype}

\usepackage{inconsolata}

\usepackage{graphicx}

\usepackage{amsmath}
\usepackage{amssymb}
\usepackage{amsfonts}
\usepackage{mathtools}
\usepackage{dsfont}
\usepackage{relsize}
\usepackage{anyfontsize}
\usepackage[table]{xcolor}
\usepackage{tabularx}
\usepackage{makecell}
\usepackage{diagbox}
\usepackage[dvipsnames]{xcolor}
\usepackage{algorithm2e}
\SetKw{Input}{Input}
\SetKw{Output}{Output}
\RestyleAlgo{ruled}

\usepackage{subfigure}

\usepackage{multirow}
\usepackage{booktabs}
\usepackage{siunitx}

\usepackage{pifont}

\usepackage{enumitem}

\usepackage{xspace}
\newcommand{\gray}[1]{\textcolor{gray}{#1}}
\newcommand{\green}[1]{\textcolor{LimeGreen}{#1}}

\newcommand{\red}[1]{\textcolor{Salmon}{#1}}

\usepackage{array}
\newcolumntype{L}[1]{>{\raggedright\let\newline\\\arraybackslash\hspace{0pt}}m{#1}}
\newcolumntype{C}[1]{>{\centering\let\newline\\\arraybackslash\hspace{0pt}}m{#1}}
\newcolumntype{R}[1]{>{\raggedleft\let\newline\\\arraybackslash\hspace{0pt}}m{#1}}

\usepackage{bm}

\DeclarePairedDelimiterX{\infdivx}[2]{(}{)}{#1\;\delimsize\|\;#2}

\newcommand{\vect}[1]{\mathbf{#1}} 
\newcommand{\matr}[1]{\mathrm{\mathbf{#1}}} 
\newcommand{\eq}[1]{Eq.~\eqref{#1}}

\title{Efficient Temporal-aware Matryoshka Adaptation\\for Temporal Information Retrieval}

\author{
    Tuan-Luc Huynh\textsuperscript{1} \enskip 
    {\bf Weiqing Wang\textsuperscript{1}} \enskip
    {\bf Thuy-Trang Vu\textsuperscript{1}} \enskip 
     \\
    {\bf Trung Le\textsuperscript{1}} \enskip 
    {\bf Dragan Ga\v{s}evi\'c\textsuperscript{2}} \enskip 
    {\bf Yuan-Fang Li\textsuperscript{1}} \enskip 
    {\bf Thanh-Toan Do\textsuperscript{1}} \\
    \textsuperscript{1} Department of Data Science \& AI, Monash University, Australia \\
    \textsuperscript{2} Department of Human Centred Computing, Monash University, Australia \\
    \texttt{\{tuan.huynh1,teresa.wang,trang.vu1},\\ \texttt{trunglm,dragan.gasevic,yuanfang.li,toan.do\}@monash.edu}
}

\begin{document}
\maketitle

\begin{abstract}
Retrievers are a key bottleneck in temporal Retrieval-Augmented Generation (RAG) systems: failing to retrieve temporally relevant context can degrade downstream generation, regardless of LLM reasoning.
We propose Temporal-aware Matryoshka Representation Learning (TMRL), an efficient method that equips retrievers with temporal-aware Matryoshka embeddings.
TMRL leverages the nested structure of Matryoshka embeddings to introduce a temporal subspace, enhancing temporal encoding while preserving general semantic representations.
Experiments show that TMRL efficiently adapts diverse standard encoder-only text embedding models, achieving competitive temporal retrieval and temporal RAG performance compared to existing Matryoshka-based adaptation and temporal retrieval methods, while enabling flexible accuracy-efficiency trade-offs.\footnote{Our code is available at \href{https://github.com/LouisDo2108/TMRL}{\texttt{https://github.com/LouisDo2108/TMRL}}.}
\end{abstract}

\section{Introduction}
\label{sec:introduction}
\begin{figure}
    \centering
    \includegraphics[width=0.9\linewidth]{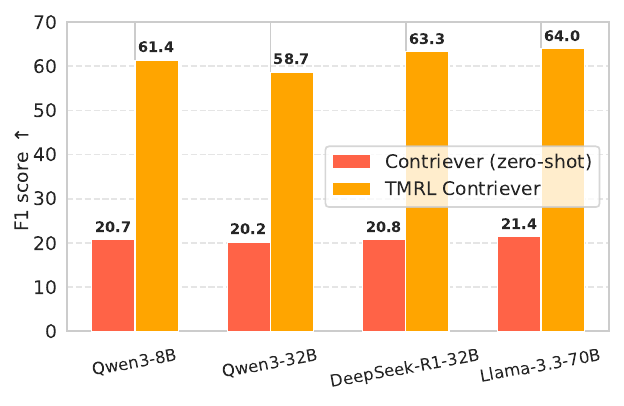}
    \caption{
    F1 score on Temporal Nobel Prize across LLMs of varying scales.
    Using larger LLMs yields negligible gains while replacing with our proposed Temporal-aware Matryoshka Representation Learning \texttt{contriever} consistently outperforms zero-shot \texttt{contriever} up to \(3\times\). This demonstrates that the 
    retriever is the primary bottleneck in temporal RAG performance.
    }
    \label{fig:1}
    \vspace{-1em}
\end{figure}

Retrieval-augmented Generation (RAG)~\cite{lewisRetrievalAugmentedGenerationKnowledgeIntensive2021} is a paradigm in information retrieval where systems generate answers grounded in external knowledge~\cite{liMatchingGenerationSurvey2025}. 
By combining LLM reasoning with retrieved evidence, RAG tackles knowledge-intensive tasks like question answering (QA), mitigating outdated or incomplete knowledge. 
Temporal QA presents unique challenges that go beyond semantic relevance. Answering time-sensitive queries requires identifying relevant time spans or even reasoning over time-evolving facts~\cite{piryaniArxiv25ItsHigh2025}. 
Driven by the strong reasoning capabilities of LLMs, recent research has proposed a myriad of datasets, analyses, and methods to benchmark Temporal QA~\cite{wang_tram_acl24,fatemiICLR25TESTTIME2025,wallat_study_acl2025,song_tempcot_sigir25}; however, they focus on LLM reasoning, overlooking the crucial role of retrieval.

We argue that in temporal RAG systems, the main bottleneck lies in the \emph{\textbf{retriever}}, which is usually a text embedding model (TEM). 
As shown in Figure~\ref{fig:1}, popular retrievers like \texttt{contriever}~\cite{izacardUnsupervisedDenseInformation2022} 
struggle with temporal retrieval, and scaling LLMs from 8B to 70B parameters cannot compensate poor retrieval.
With limited temporal encoding capability, semantic retrievers 
fail to distinguish outdated from current sources, constraining downstream generation to produce incomplete or hallucinated answers, regardless 
of the LLM's reasoning capabilities (examples in Appendix~\ref{appendix:examples_fig1}).

On the other hand, fine-tuning TEMs for temporal tasks often degrades semantic retrieval performance due to catastrophic forgetting~\cite{mccloskey_cl_1989,thrunLearningNthThing1995,han2025temporal}. Prior work~\cite{wuCIKM24TimeSensitveRetrievalAugmented2024,abdallahArxiv25TempRetrieverFusionbasedTemporal2025} mitigates this by maintaining separate temporal and semantic retrievers with query routing, 
increasing both training and inference costs.
Beyond effectiveness, temporal retrievers must also be \emph{efficient} to scale with 
growing corpora.
In this regard, Matryoshka Representation Learning (MRL)~\cite{kusupatiNeurIPS22MatryoshkaRepresentation} has emerged as a promising paradigm in modern TEMs~\cite{lee_gecko_arxiv24,nussbaumTMLR25NomicEmbed2025}, producing hierarchically structured embeddings that support flexible truncation for accuracy-efficiency trade-offs.
However, extending MRL to temporal retrieval is challenging, as temporal signals are often sparse and easily overshadowed by dominant semantic representation, particularly under aggressive embedding truncation. Moreover, naively injecting temporal supervision into Matryoshka embeddings can exacerbate the trade-off between preserving semantic retrieval quality and encoding fine-grained temporal constraints, motivating a unified and efficient temporal-aware MRL approach. 

We propose \emph{Temporal-aware Matryoshka Representation Learning (TMRL)}, an efficient adaptation method that equips TEMs with temporal-aware Matryoshka embeddings while preserving general semantic performance. 
Our approach is grounded in the insight that Matryoshka embeddings' hierarchical structure can be explicitly partitioned to preserve sparse temporal signals across dimensionalities. By dedicating a compact prefix to temporal alignment and leveraging progressive contextual expansion, TMRL enables flexible index size, lower latency yet potentially near lossless temporal retrieval via funnel retrieval.
To provide clean temporal supervision for TMRL, we refine the Temporal Nobel Prize (TNP) dataset~\cite{wuCIKM24TimeSensitveRetrievalAugmented2024} using a temporal data augmentation pipeline that generates paragraph-level temporal contrastive examples, reducing conflicts caused by multiple temporal expressions.
We develop TMRL using our augmented TNP dataset and extend evaluation to the TimeQA dataset~\cite{chenNeurIPS21DatasetAnswering2021}, demonstrating strong temporal retrieval performance, efficiency, and robustness. In summary, we make the following contributions:
\begin{itemize}[leftmargin=8pt]
\item We propose Temporal-aware Matryoshka Representation Learning (TMRL), the first efficient, plug-and-play, temporal-aware Matryoshka adaptation method for text embedding models.
\item We introduce a temporal contrastive learning augmentation pipeline that refines and augments existing temporal datasets with clean training examples for temporal information retrieval.
\item Experiments on the TNP and TimeQA datasets across six popular TEMs demonstrate TMRL’s effectiveness and efficiency for temporal retrieval and temporal RAG.
\end{itemize}

\section{Preliminary}

\paragraph{Dense Retrieval.}
TEMs learn an encoder \(f_{\theta}(\cdot)\) that maps an input sequence of \(L\) tokens \(q = (q_1, \ldots, q_L)\) into a sequence of contextualized token representations:
\(f_{\theta}(q) = [\mathbf{h}_{\texttt{[CLS]}}, \mathbf{h}_1, \ldots, \mathbf{h}_L] \in \mathbb{R}^{(L+1) \times d}\),
where \(\mathbf{h}_i \in \mathbb{R}^{d}\) denotes the hidden representation of the \(i\)-th token. 
A pooling function aggregates these token-level representations into a fixed-dimensional vector:
\(\vect{z}_q = \mathrm{Pool}(f_{\theta}(q)) \in \mathbb{R}^{d}\).

Most TEMs are optimized with a fixed \(d\)-dimensional InfoNCE loss with in-batch negatives~\citep{oord_infonce_arxiv18,zhaoDenseTextRetrieval2024}.
Given a query \(q\), a positive passage \(p^{+}\), and a set of negative passages \(\mathcal{N}_q = \{p^{-}_1, \ldots, p^{-}_n\}\), the loss is:
\begin{equation}
\begin{gathered}
\label{eq:infonce}
\mathcal{L}_{\mathrm{InfoNCE}}^{(d)}(q, p^{+}, \mathcal{N}_q)
= -\log \\
\frac{
    e^{\,\mathrm{sim}^{(d)}(\mathbf{z}_q, \mathbf{z}_{p^{+}}) / \tau}
}{
    e^{\,\mathrm{sim}^{(d)}(\mathbf{z}_q, \mathbf{z}_{p^{+}}) / \tau}
    + \sum_{i=1}^{n} e^{\,\mathrm{sim}^{(d)}(\mathbf{z}_q, \mathbf{z}_{p^{-}_i}) / \tau}
},
\end{gathered}
\end{equation}
where \(\tau\) is a temperature hyperparameter.
Let \(\vect{x}_{\leq m}\) denote the slicing of the first \(m\) dimensions of a vector \(\vect{x}\). The similarity function is typically cosine similarity:
\begin{equation}
\label{eq:similarity_function_truncated}
\mathrm{sim}^{(m)}(\vect{x}, \vect{y})
= \frac{\langle \vect{x}_{\leq m},\, \vect{y}_{\leq m} \rangle}{\lVert \vect{x}_{\leq m} \rVert \lVert \vect{y}_{\leq m} \rVert}.
\end{equation}
\paragraph{Matryoshka Representation Learning (MRL).}
MRL~\cite{kusupatiNeurIPS22MatryoshkaRepresentation} endows TEMs with nested representations for adaptive accuracy–efficiency trade-offs. Formally, given a set of target dimensions $\mathcal{M} \subset [d]$, the truncated embeddings $(\mathbf{z}_q)_{\leq m}$ for all $m \in \mathcal{M}$ remain independently functional for retrieval.
\section{Method}
\label{sec:method}
\begin{figure*}[ht]
\centering
\includegraphics[width=0.9\textwidth]{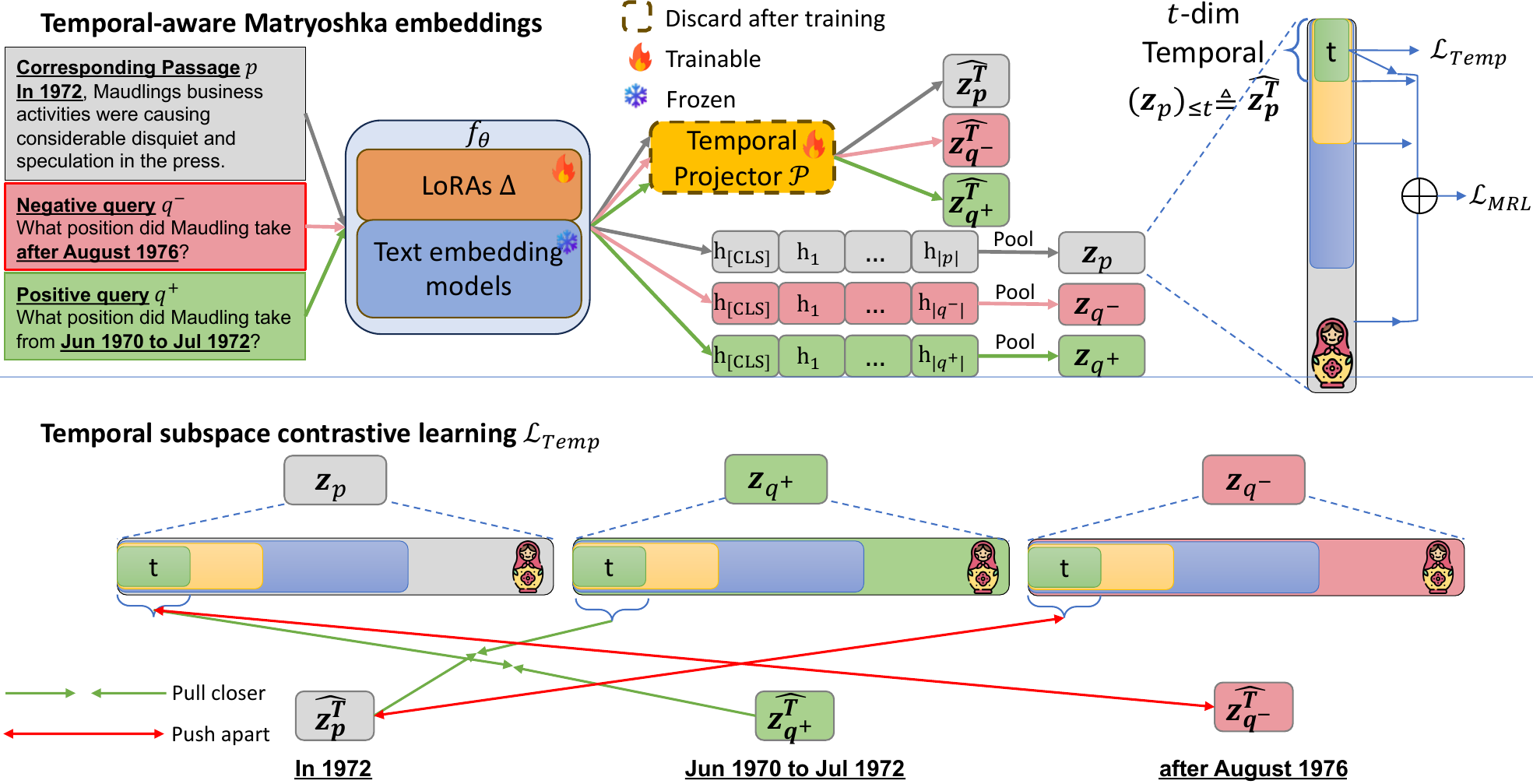}
\caption{
Overview of Temporal-aware Matryoshka Representation Learning. 
The temporal projector $\mathcal{P}$ learns to compress temporal tokens (Eq.~\ref{eq:temporal_tokens}) into a compact $t$-dimensional temporal representation (Eq.~\ref{eq:projected_temporal_tokens_mean_pooling}).
These representations are used in the temporal subspace contrastive loss $\mathcal{L}_{\mathrm{Temp}}$ (Eq.~\ref{eq:temporal_query_infonce}) to endow the Matryoshka embeddings with a dedicated temporal subspace, yielding the Temporal-aware Matryoshka embeddings (Eq.~\ref{eq:temporal_matryoshka}).
}
\label{fig:tmrl}
\end{figure*}

In this section, we perform a preliminary study to motivate Temporal-aware Matryoshka Representation Learning (TMRL), which introduces a temporal subspace to enhance temporal representations across Matryoshka embedding dimensions (Figure~\ref{fig:tmrl}).
Subsequently, we introduce the temporal contrastive learning augmentation pipeline, designed specifically for training TMRL (Figure~\ref{fig:data_comparison}).

\subsection{Preliminary Study}
We hypothesize that global pooling mechanisms in TEMs inadvertently suppress temporal signals, as they are optimized for global semantic aggregation. 
To investigate this, we conduct a preliminary analysis on the contextualized token-level representations produced by existing TEMs. 

Let \(\mathcal{T}_q \subseteq \{1, \dots, L\}\) denote the set of token indices corresponding to temporal expressions in the input query. 
To maintain compatibility with both \texttt{[CLS]}-pooling and mean-pooling TEMs,
we define a conditional set \(\mathcal{C}\) where \(\mathcal{C} = \{\texttt{[CLS]}\}\) for \texttt{[CLS]}-pooling TEMs and \(\mathcal{C} = \emptyset\) otherwise.
The contextualized temporal and semantic representations are defined as:
\begin{equation}
\label{eq:temporal_tokens}
    \vect{z}^{T}_{q} = \operatorname{MeanPool}\left(\{\, \mathbf{h}_i \ {\mid} \ i \in \mathcal{T}_q \cup \mathcal{C} \,\}\right).
\end{equation}
\begin{equation}
\resizebox{\linewidth}{!}{
\label{eq:non_temporal_tokens}
    \(\vect{z}^{S}_{q} = \operatorname{MeanPool}\left(\{\, \mathbf{h}_i \ {\mid} \ i \in (\{1,\dots,L\} \setminus \mathcal{T}_q) \cup \mathcal{C} \,\}\right).\)
}
\end{equation}
We include the \texttt{[CLS]} token in both representations since it aggregates global information and may encode temporal signals. We evaluate retrieval performance on the TNP benchmark~\citep{wuCIKM24TimeSensitveRetrievalAugmented2024} using the proposed representations.

\begin{table}[t]
\setlength{\tabcolsep}{1.0pt}
\renewcommand{\arraystretch}{1.0}
\centering
\scalebox{0.75}{
\begin{tabular}{l|cccccc}
\toprule
\multirow{2}{*}{Representation} & \texttt{contriever} & \texttt{nomic} & \texttt{gte}   & \texttt{gte1.5} & \texttt{bge}   & \texttt{bge-m3} \\
& Mean & Mean & Mean & \texttt{CLS} & \texttt{CLS} & \texttt{CLS} \\
\midrule
Joint (zero-shot) \(\mathbf{z}_q\) & 21.50 & \textbf{78.01} & \textbf{80.47} & 52.17  & 72.63 & 59.62  \\
Non-temporal \(\mathbf{z}^S_{q}\)  & 10.90 & 74.91 & 80.31 & 46.91 & \textbf{73.51} & 57.04  \\
Temporal \(\mathbf{z}^T_{q}\) & \textbf{24.47} & 71.66 & 79.60 & \textbf{52.77}  & 69.60 & \textbf{61.40} \\
\bottomrule
\end{tabular}
}
\caption{
nDCG{@}10 on TNP using different TEMs with different representations. 
The average number of semantic and temporal tokens before mean pooling is 18.5 and 3.5, respectively.}
\label{tab:temporal_non_temporal_prelim}
\end{table}

Results in Table~\ref{tab:temporal_non_temporal_prelim} demonstrate that despite being sparse (i.e., comprising 5\(\times\) fewer tokens), \(\vect{z}^{T}_{q}\) closely matches or even outperforms \(\vect{z}^{S}_{q}\) on TIR tasks. 
This reveals two key insights: 
\begin{itemize}[leftmargin=8pt]
    \item Contextualized temporal tokens constitute a highly informative, compact signal for TIR.
    \item Existing TEM embeddings fail to explicitly preserve and leverage these signals.
\end{itemize}

While Standard Matryoshka adaptation~\citep{yoonEMNLP24MatryoshkaAdaptorUnsupervised2024} yields hierarchical embeddings, it does not strengthen temporal encoding. 
Conversely, naive temporal contrastive learning enhances temporal cues but does not introduce a nested structure required for Matryoshka retrieval.
Therefore, ensuring consistent temporal alignment across all Matryoshka embedding dimensions is non-trivial.
These observations motivate a research question:

\emph{\textbf{How should we adapt TEMs with temporal-aware Matryoshka embeddings?}}

\subsection{Temporal-aware Matryoshka Representation Learning (TMRL)}
\subsubsection{Matryoshka Temporal Subspace}
Minimizing the InfoNCE loss maximizes a variational lower bound on mutual information (MI) between a pair of query and document representations~\cite{poole_mutualinformationbound_icml19}.
In standard MRL, this corresponds to maximizing a lower bound on the cumulative MI across all truncation scales
\(\sum_{m \in \mathcal{M}} I\left((\vect{z}_q)_{\leq m}; (\vect{z}_p)_{\leq m}\right)\) 
(proof in Appendix~\ref{appendix:proof}).
However, static truncation in standard MRL degrades information~\cite{zhang_smec_emnlp25}. 
In TIR, sparse temporal cues are scattered by standard contrastive alignment, making them highly vulnerable to dimension reduction. 
Standard MRL does not explicitly guarantee these temporal cues are preserved in the prefix dimensions. 
Therefore, to explicitly preserve temporal signals across all scales, we maximize the temporal MI for query-to-passage alignment:
\(I(\vect{z}_q^T; (\vect{z}_p)_{\leq m})\).
By the MI chain rule:
\begin{gather*}
I\left(\vect{z}_q^T; (\vect{z}_p)_{\leq m}\right) = \\
I\left(\vect{z}_q^T; (\vect{z}_p)_{\leq t}\right) + 
I\left(\vect{z}_q^T; (\vect{z}_p)_{(t+1):m} {\mid} (\vect{z}_p)_{\leq t}\right).
\end{gather*}
Based on this decomposition, we explicitly maximize its prefix term, \(I(\vect{z}_q^T; (\vect{z}_p)_{\leq t})\), where \(t \leq d\) denotes a compact temporal subspace dimension chosen to reflect the inherent sparsity of temporal cues.
Setting \((\vect{z}_p)_{\leq t} \triangleq \vect{z}_p^T\) guarantees temporal fidelity at the prefix and natural inheritance at all \(m > t\).
As dimensionality increases, the unconstrained suffix progressively captures additional retrieval signals beyond this fixed temporal prefix.
The above derivations and design symmetrically apply to the passage-to-query direction.

To operationalize this allocation, we introduce a temporal projector \(\mathcal{P} \colon \mathbb{R}^{d} \to \mathbb{R}^{t}\) (a two-layer FFN) 
to compress temporal tokens \(\{\mathbf{h}_i \mid i \in \mathcal{T}_q \cup \mathcal{C}\}\) into a dimensionally compatible surrogate. 
While inherently lossy, this projection preserves essential temporal signals while enabling efficient prefix allocation. The final projected temporal representation is obtained via mean-pooling:
\begin{equation}
\label{eq:projected_temporal_tokens_mean_pooling}
    \widehat{\vect{z}_q^{T}}
    =
    \operatorname{MeanPool}\left(\{\, \mathcal{P}(\mathbf{h}_i) \ {\mid} \ i \in \mathcal{T}_q \cup \mathcal{C} \,\}\right).
\end{equation}
This defines the temporal prefix, yielding a structured representation where each truncated embedding decomposes into a fixed temporal subspace and a progressively expanding contextual suffix:
\begin{gather}
    \label{eq:temporal_subspace}
    (\vect{z}_{q})_{\leq t} \coloneqq \widehat{\vect{z}_q^{T}}. \\
    \label{eq:temporal_matryoshka}
    (\vect{z}_{q})_{\leq m}
    \coloneqq
    \left[\, \widehat{\vect{z}_q^{T}} \ \Vert \ (\vect{z}_{q})_{(t+1):m} \,\right].
\end{gather}

\subsubsection{Temporal Subspace Contrastive Learning}
\label{subsubsec:temporal_mrl}
Following~\citet{wuCIKM24TimeSensitveRetrievalAugmented2024}, temporal contrastive learning operates on positive and negative queries rather than passages, with the negative set defined as \(\mathcal{N}_p=\{q^{-}_1,\ldots,q^{-}_n\}\).
We optimize the temporal subspace via a dedicated contrastive loss (bottom of Figure~\ref{fig:tmrl}).
The InfoNCE loss for the $t$-dimensional temporal subspace is defined as:
\begin{equation}
\begin{gathered}
\label{eq:temporal_query_infonce}
\mathcal{L}^{\mathrm{q}}_{\mathrm{Temp}}(p, q^{+}, \mathcal{N}_p)
= -\log \\
\frac{
    e^{\,\mathrm{sim}^{(t)}\left(\gray{\widehat{\vect{z}^{T}_p}},\, \green{\vect{z}_{q^+}}\right) / \tau}
}{
    e^{\,\mathrm{sim}^{(t)}\left(\gray{\widehat{\vect{z}^{T}_p}},\, \green{\vect{z}_{q^+}}\right) / \tau}
    +
    \sum\limits_{i=1}^{n}
    e^{\,\mathrm{sim}^{(t)}\left(\gray{\widehat{\vect{z}^{T}_p}},\, \red{\vect{z}_{q^{-}_i}}\right) / \tau}
},\\
\mathcal{L}^{\mathrm{p}}_{\mathrm{Temp}}(p, q^{+}, \mathcal{N}_p)
= -\log \\
\frac{
    e^{\,\mathrm{sim}^{(t)}\left(\gray{\vect{z}_p},\, \green{\widehat{\vect{z}^T_{q^+}}}\right) / \tau}
}{
    e^{\,\mathrm{sim}^{(t)}\left(\gray{\vect{z}_p},\, \green{\widehat{\vect{z}^T_{q^+}}}\right) / \tau}
    +
    \sum\limits_{i=1}^{n}
    e^{\,\mathrm{sim}^{(t)}\left(\gray{\vect{z}_p},\, \red{\widehat{\vect{z}^T_{q^-_i}}}\right) / \tau}
},\\
\mathcal{L}_{\mathrm{Temp}} = 
\mathcal{L}^{\mathrm{q}}_{\mathrm{Temp}} + \mathcal{L}^{\mathrm{p}}_{\mathrm{Temp}},
\end{gathered}
\end{equation}
where \gray{\(\widehat{\vect{z}^{T}_p}\)}, \green{\(\widehat{\vect{z}^T_{q^+}}\)}, 
and \red{\(\widehat{\vect{z}^T_{q^-_i}}\)} denote the passage, positive query, and negative query counterparts of \eq{eq:temporal_tokens}.

\subsubsection{Local and Structural Self-distillation}
We frame MRL as self-distillation, where higher-dimensional embeddings supervise truncated counterparts~\cite{yoonEMNLP24MatryoshkaAdaptorUnsupervised2024}. 
To enhance temporal-aware Matryoshka representations, we introduce an auxiliary objective that preserves local structure by aligning teacher–student nearest-neighbor similarities. 
Given a batch of embeddings \(\matr{X}^{(d)} \in \mathbb{R}^{B \times d}\) and truncated counterparts \(\matr{X}^{(m)} \in \mathbb{R}^{B \times m}\) (\(d > m\)), we apply a top-\(k\) pairwise similarity loss:
\begin{equation}
\small
\label{eq:loss_pairwise_topk}
\mathcal{L}_{\mathrm{Dist}}
=
\sum\limits_{\mathclap{\substack{m\in\mathcal{M}, \\ m\neq d}}}
\sum_i
\sum_{j \in \mathcal{N}}
\left\lVert
\mathrm{sim}^{(d)}(\matr{x}_i, \matr{x}_j)
-
\mathrm{sim}^{(m)}(\matr{x}_i, \matr{x}_j)
\right\rVert_1,
\end{equation}
where \(\mathcal{N} {=} \operatorname{arg\,top}_k \mathrm{sim}^{(d)}(\matr{x}_i, \matr{x}_j)\). To complement local pairwise consistency, we adopt linear Centered Kernel Alignment (CKA)~\cite{kornblith2019similarity} as a global structural regularizer, ensuring truncated embeddings preserve the teacher's overall representational geometry.

The linear CKA between two embedding sets can be expressed using covariance \(\mathrm{cov(\cdot,\cdot)}\) as:
\begin{equation}
\mathrm{CKA}(\matr{X}, \matr{Y}) = \frac{\| \mathrm{cov}(\matr{X}, \matr{Y}) \|_F^2}{\| \mathrm{cov}(\matr{X}, \matr{X}) \|_F \, \| \mathrm{cov}(\matr{Y}, \matr{Y}) \|_F}.
\end{equation}
Let us denote the centered variants of the \(d\) and \(m\)-dimensional embeddings as:
\begin{equation}
\begin{gathered}
\tilde{\matr{X}}^{(d)} = \matr{X}^{(d)} - \frac{1}{B} \mathbf{1}\mathbf{1}^\top \matr{X}^{(d)}.
\\
\tilde{\matr{X}}^{(m)} = \matr{X}^{(m)} - \frac{1}{B} \mathbf{1}\mathbf{1}^\top \matr{X}^{(m)}.
\end{gathered}
\end{equation}
The auxiliary Matryoshka CKA is defined as:
\begin{equation}
\mathcal{L}_{\mathrm{CKA}}
=
\sum\limits_{\mathclap{\substack{m\in\mathcal{M}, m\neq d}}}(1 - \mathrm{CKA}(\tilde{\matr{X}}^{(d)}, \tilde{\matr{X}}^{(m)}))
\end{equation}

\subsubsection{Optimization Objective}
\label{par:training_objectives}
TMRL-based TEMs are fine-tuned using LoRAs~\citep{huLoRALowRankAdaptation2021} to minimize catastrophic forgetting and enable zero-latency inference via adapter merging.
The optimization objective is:
\begin{equation}
\begin{gathered}
\label{eq:final_loss}
\mathcal{L}_{\mathrm{TMRL}} = \mathcal{L}_{\mathrm{MRL}} 
{+} \alpha\mathcal{L}_{\mathrm{Temp}}
{+} \beta\mathcal{L}_{\mathrm{Dist}} 
{+} \gamma\mathcal{L}_{\mathrm{CKA}}.
\end{gathered}
\end{equation}
where \(\alpha,\beta,\gamma\) are hyperparameters (ablation studies in Section~\ref{subsec:discussions}) and \(\mathcal{L}_{\mathrm{MRL}} = \sum_{m \in \mathcal{M}} \mathcal{L}_{\mathrm{InfoNCE}}^{(m)}\).

\subsection{Temporal Contrastive Learning Augmentation}
\label{sec:data}
\begin{figure*}[ht]
\centering
\includegraphics[width=0.8\textwidth]{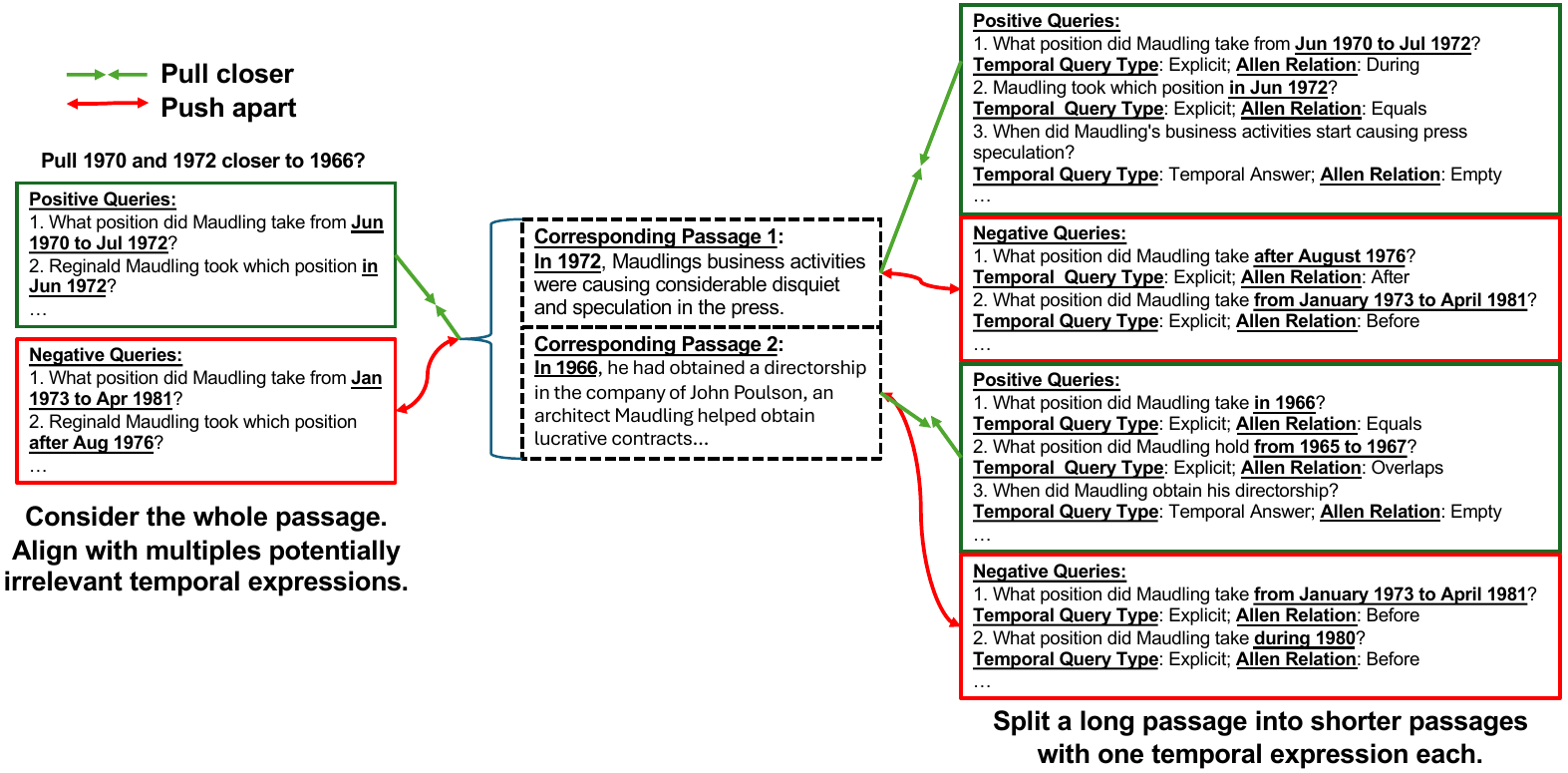}
\caption{
Comparison between the TNP's training dataset (left) and our augmented version (right).
We follow a simple principle: split a long passage into shorter passages, discard sentences with multiple temporal expressions, ensuring that each passage contains one temporal expression before generating augmented positive/negative queries.
}
\label{fig:data_comparison}
\end{figure*}

The temporal projector \(\mathcal{P}\) in TMRL requires clean, unambiguous temporal expressions. However, the Temporal Nobel Prize (TNP) dataset~\cite{wuCIKM24TimeSensitveRetrievalAugmented2024} often contains passages with multiple temporal expressions (e.g., ``1972'' and ``1966''), which introduce conflicting gradient signals during contrastive training (Figure~\ref{fig:data_comparison}, left). To address this, we refine the \emph{\textbf{train dataset}} via two steps:

\paragraph{Passage Splitting.} 
\label{par:passage_splitting}
Using SUTime~\cite{chang_sutime_2012}, we identify temporal expressions at the sentence level and iteratively merge adjacent sentences until a second expression would be introduced. Sentences containing multiple expressions are discarded. This yields passages with exactly one temporal expression while preserving maximal local context.

\paragraph{Positive/Negative Query Augmentation.}
\label{par:positive_negative_query_augmentation}
We generate contrastive query-passage pairs by filtering and augmenting along three dimensions: 
\begin{enumerate}[topsep=0pt, itemsep=0pt, leftmargin=12pt]
    \item We target L2 (time--event) and L3 (event--event) understanding~\cite{tanACL23BenchmarkingImproving2023}, excluding L1 (time--time) arithmetic and ordinal queries~\cite{jia2021complexcikm21} that lack reliable corpus anchors.
    \item We include temporal questions with \emph{explicit}, \emph{implicit}, and \emph{temporal answer}~\cite{jia2021complexcikm21}.
    \item We apply Allen’s interval relations~\cite{allen1983maintaining} to explicitly control query--passage temporal alignment.
\end{enumerate}

Details about the augmentation pipeline are provided in Appendix~\ref{appendix:augmentation_pipeline}.
We validate the effectiveness of this pipeline in Section~\ref{subsec:augmented_data_result_tnp}.
It is worth noting that models are evaluated on \emph{\textbf{test datasets}} consisting of unfiltered real-world documents, which serve as stress tests for model's robustness on passages with multiple temporal expressions.

\section{Experiments}
\label{sec:experiments}

\subsection{Experimental Setup}
\label{subsec:baselines}

\paragraph{Datasets and Metrics.}
We evaluate on Temporal Nobel Prize (TNP)~\cite{wuCIKM24TimeSensitveRetrievalAugmented2024} and extend to TimeQA~\cite{chenNeurIPS21DatasetAnswering2021}.
TNP decouples temporal retrieval from semantic matching, cleanly attributing gains to temporal modeling. In contrast, TimeQA leverages Wikipedia's dynamic entity histories to provide a temporally realistic benchmark.
Refer to Appendix~\ref{appendix:dataset_stats} for further details about the datasets.
Following BEIR~\cite{thakurBEIRHeterogeneousBenchmark}, we report nDCG{@}10 and Recall{@}100.
To assess semantic retrieval performance, we report nDCG{@}10 on the Natural Questions (NQ) dataset~\cite{kwiatkowskiNaturalQuestionsBenchmark2019} from BEIR using the original embedding dimension of each TEM.
We further evaluate on downstream Temporal RAG, using F1 score as the evaluation metric.
We focus on evaluating temporal retrieval, while multi-hop reasoning is handled downstream by an LLM.
Refer to Appendix~\ref{appendix:llm_settings_for_temporal_rag} for further details about  Temporal RAG setup.

\paragraph{Baselines.}
We categorize our baselines into two groups based on the evaluation objective:

\begin{itemize}[nosep, leftmargin=12pt]
    \item \textbf{Temporal Retrieval.} 
    To assess temporal encoding in full-dimensional embeddings, we compare TMRL against: 
    (i) zero-shot performance of original TEMs; 
    (ii) BM25~\cite{robertsonProbabilisticRelevanceFramework2009}; and 
    (iii) prior dense temporal retrieval methods, including Ts-Retriever~\cite{wuCIKM24TimeSensitveRetrievalAugmented2024} and TempRetriever~\cite{abdallahArxiv25TempRetrieverFusionbasedTemporal2025}.
    \item \textbf{Matryoshka Adaptation.} 
    To evaluate our proposed Matryoshka adaptation strategy, we compare against parameter-efficient fine-tuning adapatation approaches: 
    Matryoshka Adaptor (M-Adaptor)~\cite{yoonEMNLP24MatryoshkaAdaptorUnsupervised2024} and a vanilla LoRA-based MRL baseline.
\end{itemize}

\noindent\textbf{Backbones.} 
We evaluate all methods across six open-source encoder-only TEMs (\({<}\)600M parameters):
\texttt{contriever}~\cite{izacardUnsupervisedDenseInformation2022}, 
\texttt{gte-base} (\texttt{gte})~\cite{liArxiv23GeneralText2023}, 
\texttt{nomic-embed-text-v1.5}(\texttt{nomic})~\cite{nussbaumTMLR25NomicEmbed2025}, 
\texttt{gte-base-en-v1.5} (\texttt{gte1.5})~\cite{zhang_mgte_emnlp24}, 
\texttt{bge-base-en-v1.5} (\texttt{bge})~\cite{xiaoSIGIR24CPackPacked2024}, 
\texttt{bge-m3}~\cite{chenArxiv24BGEM3Embedding2024}.
Implementation details are provided in Appendix~\ref{appendix:implementation_details}.

\begin{table*}[ht]
\renewcommand{\arraystretch}{1.0}
\centering
\scalebox{0.6}{
\begin{tabular}{c|c|ccccc|ccccc|ccccc}

\toprule
\multirow{2}{*}{\textbf{Backbone}}
& \multirowcell{2}{\textbf{Train}\\\textbf{set}}
& \multicolumn{5}{c|}{\textbf{nDCG@10}}
& \multicolumn{5}{c|}{\textbf{Recall@100}}
& \multicolumn{5}{c}{\textbf{BEIR NQ nDCG@10}} \\
\cmidrule(lr){3-7} \cmidrule(lr){8-12} \cmidrule(lr){13-17}
& & 0-shot & TsR & TempR & MA & LoRA & 0-shot & TsR & TempR & MA & LoRA & 0-shot & TsR & TempR & MA & LoRA \\

\midrule
\texttt{BM25} & - & 34.58 & - & - & - & - & 76.61 & - & - & - & - & 
32.90 & - & - & - & - \\

\midrule
\multirowcell{2}{
\texttt{contriever}\\
(Mean-768)
}
& Original
& \multirow{2}{*}{21.50} & 53.15 & 57.93 & 19.75 & 56.91 
& \multirow{2}{*}{85.14} & 95.46 & 95.23 & 81.62 & \underline{96.37} 
& \multirow{2}{*}{\textbf{25.37}} & 14.60 & 19.94 & 17.63 & 18.80 
\\
& Our 
& & \cellcolor{green!25}53.64 & \cellcolor{green!25}\textbf{63.01} & \cellcolor{green!25}20.90 & \cellcolor{green!25}\underline{60.39} 
& & \cellcolor{green!25}95.54 & \cellcolor{green!25}96.21 & \cellcolor{green!25}84.24 & \cellcolor{green!25}\textbf{97.08} 
& & \cellcolor{green!25}15.22 & \cellcolor{green!25}\underline{21.42} & \cellcolor{green!25}17.68 & \cellcolor{green!25}19.46 
\\

\midrule
\multirowcell{2}{
\texttt{gte}\\
(Mean-768)
}
& Original
& \multirow{2}{*}{\textbf{80.47}} & -\(^{*}\) & -\(^{*}\) & 78.65 & 76.04 
& \multirow{2}{*}{\textbf{97.56}} & -\(^{*}\) & -\(^{*}\) & 97.50 & 96.24 
& \multirow{2}{*}{\textbf{52.84}} & -\(^{*}\) & -\(^{*}\) & 52.44 & 44.58 
\\
& Our 
& & -\(^{*}\) & -\(^{*}\) & \cellcolor{green!25}78.72 & \cellcolor{green!25}\underline{80.34} 
& & -\(^{*}\) & -\(^{*}\) & \cellcolor{green!25}\underline{97.52} & \cellcolor{green!25}97.13 
& & -\(^{*}\) & -\(^{*}\) & \cellcolor{green!25}\underline{52.53} & \cellcolor{green!25}50.78 
\\

\midrule
\multirowcell{2}{
\texttt{nomic}\\
(Mean-768)
}
& Original
& \multirow{2}{*}{78.01} & 56.25 & -\(^{*}\) & 75.10 & \underline{81.59} 
& \multirow{2}{*}{95.60} & 92.22 & -\(^{*}\) & \cellcolor{green!25}95.79 & \underline{96.89} 
& \multirow{2}{*}{59.73} & 36.66 & -\(^{*}\) & 58.88 & \underline{62.50} 
\\
& Our 
& & \cellcolor{green!25}60.53 & -\(^{*}\) & \cellcolor{green!25}75.14 & \cellcolor{green!25}\textbf{82.55} 
& & \cellcolor{green!25}93.87 & -\(^{*}\) & 95.75 & \cellcolor{green!25}\textbf{97.58} 
& & \cellcolor{green!25}41.13 & -\(^{*}\) & \cellcolor{green!25}58.94 & \cellcolor{green!25}\textbf{62.76} 
\\

\midrule
\multirowcell{2}{
\texttt{bge}\\
(\texttt{CLS}-768)
}
& Original
& \multirow{2}{*}{72.63} & 44.75 & -\(^{*}\) & 69.80 & \cellcolor{green!25}\textbf{78.00} 
& \multirow{2}{*}{\textbf{97.87}} & 89.47 & -\(^{*}\) & 97.38 & 97.49 
& \multirow{2}{*}{\textbf{54.15}} & \cellcolor{green!25}37.22 & -\(^{*}\) & 53.01 & 50.97 
\\
& Our 
& & \cellcolor{green!25}59.87 & -\(^{*}\) & \cellcolor{green!25}70.14 & \underline{77.42} 
& & \cellcolor{green!25}94.27 & -\(^{*}\) & \cellcolor{green!25}97.46 & \cellcolor{green!25}\underline{97.58} 
& & 34.30 & -\(^{*}\) & \cellcolor{green!25}\underline{53.14} & \cellcolor{green!25}52.47 
\\

\midrule
\multirowcell{2}{
\texttt{gte1.5}\\
(\texttt{CLS}-768)
}
& Original
& \multirow{2}{*}{52.17} & 53.14 & -\(^{*}\) & 48.65 & \underline{72.90} 
& \multirow{2}{*}{94.71} & 92.94 & -\(^{*}\) & 93.84 & \cellcolor{green!25}\textbf{96.42} 
& \multirow{2}{*}{\textbf{52.96}} & \cellcolor{green!25}41.81 & -\(^{*}\) & 52.70 & 52.03 
\\
& Our 
& & \cellcolor{green!25}67.10 & -\(^{*}\) & \cellcolor{green!25}50.29 & \cellcolor{green!25}\textbf{74.00} 
& & \cellcolor{green!25}95.54 & -\(^{*}\) & \cellcolor{green!25}94.20 & \underline{96.09} 
& & 40.86 & -\(^{*}\) & \cellcolor{green!25}52.75 & \cellcolor{green!25}\underline{52.78} 
\\

\midrule
\multirowcell{2}{
\texttt{bge-m3}\\
(\texttt{CLS}-1024)
}
& Original
& \multirow{2}{*}{59.62} & 43.84 & -\(^{*}\) & 57.65 & \underline{78.51} 
& \multirow{2}{*}{\textbf{96.99}} & 88.57 & -\(^{*}\) & 96.55 & 96.27 
& \multirow{2}{*}{\textbf{60.57}} & 18.56 & -\(^{*}\) & 59.92 & 57.81 
\\
& Our 
& & \cellcolor{green!25}55.31 & -\(^{*}\) & \cellcolor{green!25}57.83 & \cellcolor{green!25}\textbf{80.23} 
& & \cellcolor{green!25}95.36 & -\(^{*}\) & \cellcolor{green!25}96.56 & \cellcolor{green!25}\underline{96.47} 
& & \cellcolor{green!25}23.35 & -\(^{*}\) & \cellcolor{green!25}\underline{59.96} & \cellcolor{green!25}59.54 
\\

\bottomrule
\end{tabular}
}
\caption{
Retrieval performance on TNP (full-dimensional embeddings). 
``-'' indicates N/A; 
$^{*}$ denotes unstable training. 
Abbreviations: 0-shot (Zero-shot), TsR (Ts-Retriever), TempR (TempRetriever), MA (M-Adaptor), LoRA. ``Original'' / ``Our'' denote original/augmented training sets. \textcolor{green}{Green} highlights the superior training set per method. \textbf{Bold} and \underline{underlined} values indicate best and second-best results per TEM.
}
\label{tab:main_result_tnp_data}
\end{table*}

\begin{figure*}[!ht]
    \centering
    \includegraphics[width=0.9\linewidth]{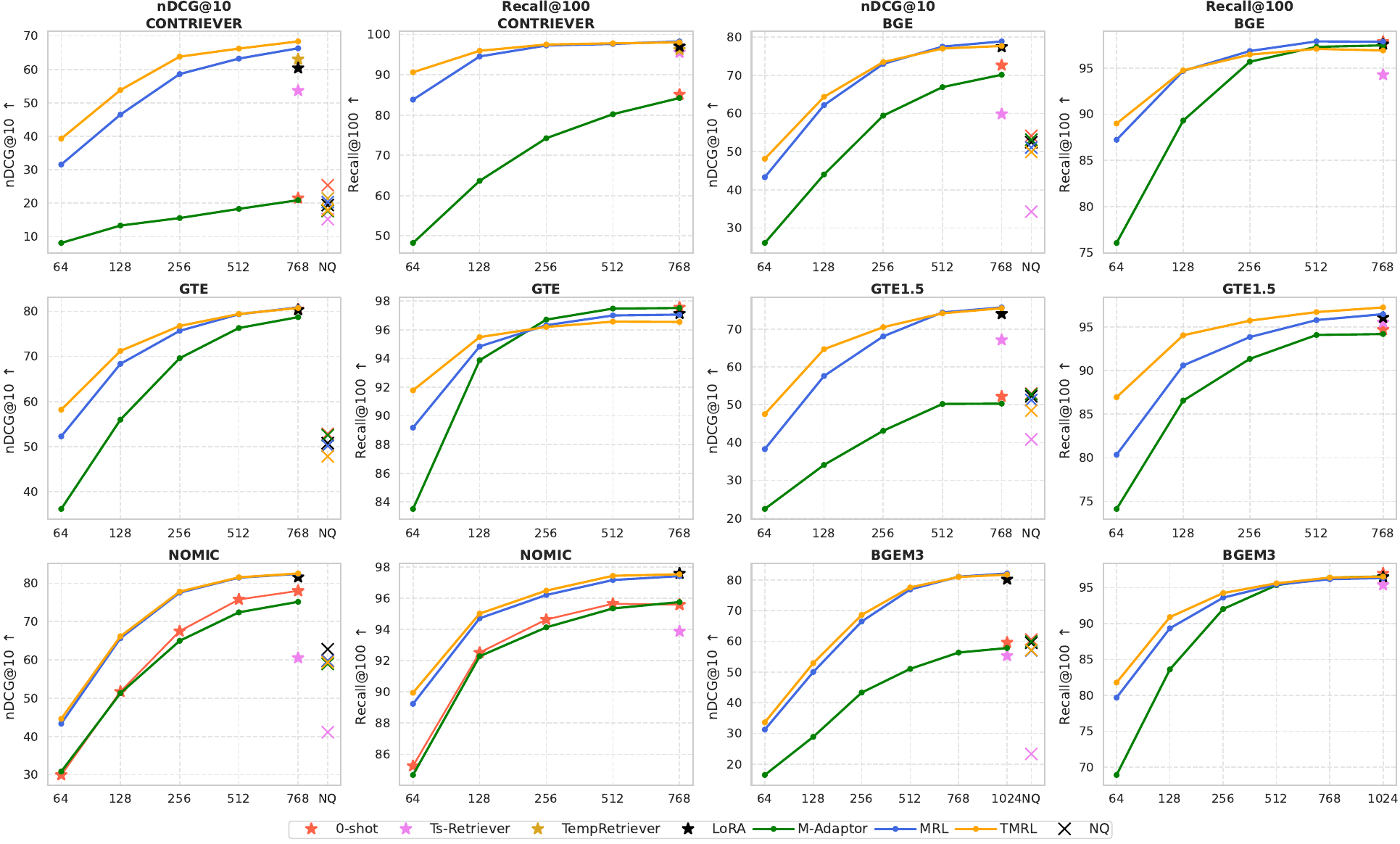}
    \caption{Retrieval performance on TNP. TMRL retrievers improves temporal retrieval, particularly nDCG@10 across embedding dimensionalities, while preserving general semantic performance on NQ.
    }
\label{fig:matryoshka_main_result_tnp}
\end{figure*}

\subsection{Validating The Augmented TNP Dataset}
\label{subsec:augmented_data_result_tnp}
Table~\ref{tab:main_result_tnp_data} demonstrates that our augmented TNP dataset yields consistent gains across methods and TEMs by providing cleaner temporal supervision. 
The results highlight limitations in prior approaches: TempRetriever peaks on \texttt{contriever} but lacks cross-TEM generalization, whereas Ts-Retriever shows stable improvements in nDCG{@}10 and Recall{@}100. 
Among parameter-efficient methods, M-Adaptor yields negligible gains, while LoRA best leverages the augmented data, consistently outperforming baselines with only marginal NQ degradation. 
Additional augmentation pipeline's ablation studies are provided in Appendix~\ref{appendix:augmentation_pipeline}.

To understand how different temporal relations (e.g., Before/After, Contains/During) influence TMRL's performance, we perform a comprehensive ablation study that systematically removes specific Allen relations from the training data in Appendix~\ref{appendix_ablation_study_allen_relations}
This serves as a quantitative proxy diagnostic since both TNP and TimeQA lack explicit Allen relation annotations for test queries.
Unless noted otherwise, all subsequent experiments use the augmented training datasets.

\subsection{Main Results}
Figures~\ref{fig:matryoshka_main_result_tnp} and~\ref{fig:matryoshka_result_timeqa} report retrieval performance on TNP and TimeQA. Across both benchmarks, M-Adaptor's performance is bounded by the TEM's zero-shot capability, suggesting limitations in semantic-only Matryoshka adaptation. \texttt{nomic}'s zero-shot degradation further indicates that temporal retrieval may be more sensitive to dimensionality reduction than semantic retrieval.

\paragraph{TNP.}
TMRL shows consistent improvements over M-Adaptor across dimensions and TEMs. 
Gains are most pronounced at lower dimensions, with nDCG{@}10 improvements suggesting that the dedicated temporal subspace helps preserve temporal signals under truncation. 
Unlike Ts-Retriever, TMRL maintains competitive semantic retrieval on NQ. 
We observe a slight Recall{@}100 drop at higher dimensions for \texttt{gte} and \texttt{bge}; however, since broader recall does not necessarily improve downstream RAG~\cite{jin_longcontextrag_iclr25}, we prioritize nDCG{@}10, which better reflects top-ranked temporal relevance for RAG pipelines.

\paragraph{TimeQA.} 
Due to space constraints, we place TimeQA results in Appendix~\ref{appendix:timeqa_main_results}.
On TimeQA, TMRL's improvements emerge at higher dimensions (i.e., \({\geq} 256\)), suggesting this benchmark relies more heavily on semantic capacity. 
Beyond 256 dimensions, TMRL matches the MRL baseline's Recall{@}100 while showing nDCG{@}10 gains, with NQ semantic performance remaining comparable. 
Our augmentation pipeline yields additional nDCG{@}10 improvements across most TEMs, with modest Recall{@}100 trade-offs at smaller scales. 
These results indicate that TMRL's temporal-aware Matryoshka embeddings can improve temporal retrieval while supporting flexible accuracy--efficiency trade-offs.

\subsection{Temporal RAG Results}
\begin{figure}[ht]
    \centering
    \includegraphics[width=0.9\linewidth]{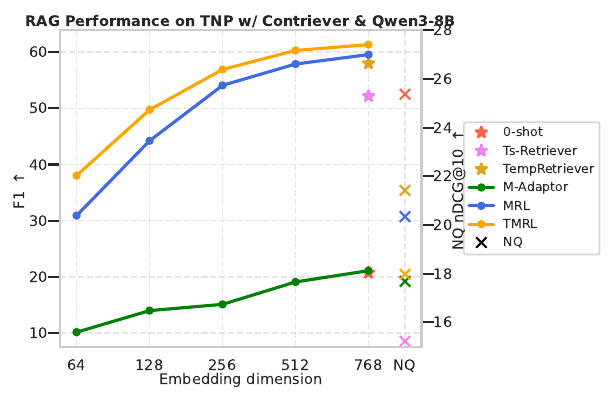}
    \caption{
    TMRL \texttt{contriever} enables the Temporal RAG system to achieve competitive TQA performance while enabling flexible accuracy-efficiency trade-off.}
    \label{fig:rag_tnp}
\end{figure}

Figures~\ref{fig:rag_tnp} and~\ref{fig:rag_timeqa} report temporal RAG F1 on TNP and TimeQA using \texttt{Qwen3-8B}~\cite{qwen3} with \texttt{contriever} variants. RAG trends closely follow the corresponding retrieval results (Figures~\ref{fig:matryoshka_main_result_tnp} and~\ref{fig:matryoshka_result_timeqa}), suggesting retriever quality is a key factor in temporal RAG performance. Across both datasets, TMRL \texttt{contriever} achieves a strong accuracy--efficiency trade-off among Matryoshka baselines. With a dedicated temporal subspace, TMRL shows improvements over other adaptation methods across dimensions, with gains most pronounced under aggressive truncation. On TNP, for instance, 64-dimensional embeddings yield over $3\times$ the F1 score of M-Adaptor; 256-dimensional embeddings match TempRetriever's performance while using only 33\% of the index size.

\subsection{Discussions}
\label{subsec:discussions}
\paragraph{Efficiency Analysis.}
Table~\ref{tab:efficiency_comparison} compares TMRL against prior methods using the \texttt{contriever} backbone. TMRL matches Ts-Retriever in storage (0.8 GiB) and inference cost (single model) but offers three key advantages: (i) parameter-efficient LoRA adaptation, (ii) stronger temporal retrieval with competitive semantic performance on NQ, and (iii) native Matryoshka support for flexible truncation. Conversely, TempRetriever uses dual bi-encoders and an external extractor (SUTime~\cite{chang_sutime_2012}), doubling storage to 1.6 GiB, increasing latency, and requiring 82 seconds to index just 1000 documents. TSM~\cite{han2025temporal} trains seven time-specific models (5.7 GiB storage), yielding highly inefficient training despite inference-time parameter averaging.

Unlike these fixed-size baselines, TMRL enables dynamic index sizing. Table~\ref{tab:tmrl_contriever_embedding_size} shows that a 256-dimensional TMRL index uses only 33\% of the full storage while retaining 92\% of nDCG{@}10 and 84\% of retrieval time. Since latency metrics cover search operations only, actual gains from reduced memory I/O and cache effects are likely higher.

\begin{table}[t]
\setlength{\tabcolsep}{1.0pt}
\renewcommand{\arraystretch}{1.0}
\centering
\scalebox{0.75}{
\begin{tabular}{l|ccccc}
\toprule
\multirow{2}{*}{Method} & Train $\downarrow$ & Inference $\downarrow$ & Storage $\downarrow$ &  Corpus Index & TNP $\uparrow$ \\
& (Model) & (Model) & (GiB) & Time $\downarrow$ (s) & (nDCG@10) \\
\midrule
TsR & 1 &  1 &  \textbf{0.8 } & \textbf{< 1} & 53.6 \\
TempR & 2$^\dagger$ & 2$^\dagger$ & 1.6 & 82 & 63.0 \\
TSM & 7 & 1 &  5.7  & \textbf{< 1} & 64.6 \\
\midrule
\textbf{TMRL} & LoRA & 1 & \textbf{0.8 } & \textbf{< 1} & \textbf{68.4} \\
\bottomrule
\end{tabular}
}
\caption{
Efficiency comparison of TMRL and prior temporal retrieval methods using \texttt{contriever} on TNP.
$^\dagger$ denotes further requiring a temporal extractor. TsR and TempR denote Ts-Retriever and TempRetriever, respectively.
}
\label{tab:efficiency_comparison}
\end{table}
\begin{table}[ht]
\renewcommand{\arraystretch}{1.0}
\centering
\scalebox{0.8}{
\begin{tabular}{l|ccccc}
\toprule
Criterion & 64 & 128 & 256 & 512 & 768 \\
\midrule
nDCG{@}10 & 26.9  & 39.8  & 45.0 & 48.1 & 49.0 \\
Ratio & 55\% & 81\% & 92\% & 98\% & 100\% \\
\midrule
Corpus Index (MiB) & 28 & 54 & 107 & 214 & 321 \\
Ratio & 9\% & 17\% & 33\% & 67\% & 100\% \\
\midrule
Latency (ms)  & 183  & 187 & 192  & 210  & 230  \\
Ratio & 79\% & 81\% & 84\% & 92\% & 100\% \\
\bottomrule
\end{tabular}
}
\caption{Performance, corpus index size, and retrieval latency (millisecond) of TMRL \texttt{contriever}'s Matryoshka embeddings on TimeQA. Retrieval latency reflects search operations only (index loading excluded).}
\label{tab:tmrl_contriever_embedding_size}
\end{table}

\paragraph{Funnel Retrieval.}
\begin{table}[ht]
\setlength{\tabcolsep}{1.0pt}
\renewcommand{\arraystretch}{1.0}
\centering
\scalebox{0.7}{
\begin{tabular}{l|cc|cc}
\toprule
\multirow{2}{*}{Method} & \multicolumn{2}{c|}{TNP} & \multicolumn{2}{c}{TimeQA} \\
\cmidrule{2-3} \cmidrule{4-5}
& nDCG@10 \(\uparrow\) & Latency (s) \(\downarrow\) & nDCG@10 \(\uparrow\) & Latency (s) \(\downarrow\) \\
\midrule
MRL           & 66.31         & \underline{0.101}   & 42.02           & \underline{0.226}   \\
TMRL          & \textbf{68.36} & \underline{0.101}  & \underline{49.00}              & \underline{0.226}   \\
\midrule
Funnel MRL    & 66.19 & \textbf{0.083}   & 39.59            & \textbf{0.185}    \\
Funnel TMRL   & \underline{68.35} & \textbf{0.083}   & 46.42            & \textbf{0.185}   \\
\midrule
\multicolumn{5}{c}{Latency (s) on 10\(\times\) larger TimeQA corpus \(\downarrow\)} \\
\midrule
Standard & \multicolumn{4}{c}{0.688} \\
Funnel & \multicolumn{4}{c}{\textbf{0.250}} \\
\bottomrule
\end{tabular}
}
\caption{Funnel retrieval (top-k=100, d=64) \(\rightarrow\) (top-k=10, d=768) vs. standard full-dimensional retrieval.}
\label{tab:funnel_retrieval}
\end{table}
Funnel retrieval leverages low-dimensional embeddings for initial candidate retrieval (e.g., $k{=}100, d{=}64$) followed by higher-dimensional re-ranking (e.g., $k{=}10, d{=}768$). Table~\ref{tab:funnel_retrieval} shows that TMRL \texttt{contriever}'s funnel pipeline on TNP maintains comparable nDCG{@}10 to full-dimensional retrieval and retains $\approx$ 95\% performance on TimeQA while reducing latency. In a scaled setting (e.g., TimeQA corpus duplicated 10$\times$ to 1.1M passages), this approach cuts retrieval time to $\approx$ 36\% of the full-dimensional baseline.

\paragraph{Effect of Temporal Subspace.}
\label{par:effect_of_temporal_subspace}
\begin{figure}[ht]
\centering
\includegraphics[width=0.45\textwidth]{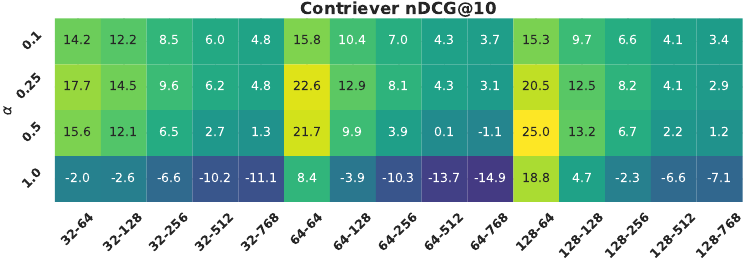}
\includegraphics[width=0.45\textwidth]{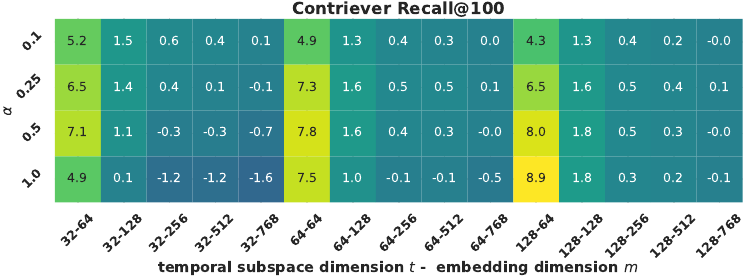}
\includegraphics[width=0.45\textwidth] {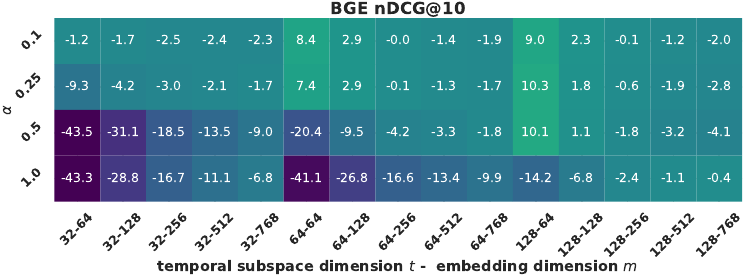} 
\includegraphics[width=0.45\textwidth]{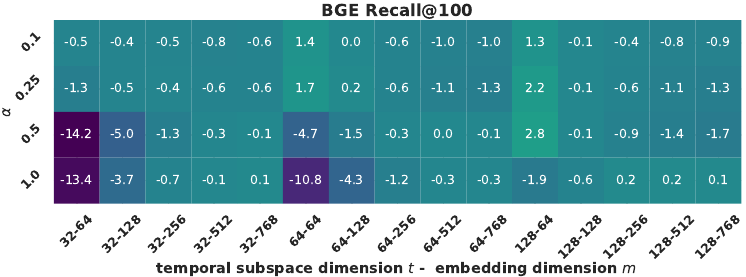}
\caption{
Ablation of the temporal subspace dimension \(t\) and the temporal subspace coefficient \(\alpha\). The results show the increase/decrease relative to the LoRA-based MRL baseline on TNP using \texttt{contriever} and \texttt{bge}.
}
\label{fig:ablation_t_alpha}
\end{figure}

\begin{table}[ht]
\setlength{\tabcolsep}{1.5pt}
\renewcommand{\arraystretch}{1.0}
\centering
\scalebox{0.8}{
\begin{tabular}{l|cccc|cccc}
\toprule
& \multicolumn{4}{c|}{\texttt{contriever}} & \multicolumn{4}{c}{\texttt{BGE}} \\
\midrule
\(\alpha\) & 0.1 & 0.25 & 0.5 & 1.0 & 0.1 & 0.25 & 0.5 & 1.0 \\
\midrule
\(t{=}32\)  & \textbf{19.67} & 18.72 & 17.04 & 15.00 & 48.81 & 47.21 & 48.71 & 45.82 \\
\(t{=}64\)  & 19.64 & 18.56 & 16.45 & 13.42 & \textbf{53.76} & 49.18 & \underline{50.62} & 49.33 \\
\(t{=}128\) & \underline{19.65} & 18.92 & 17.62 & 14.97 & 50.29 & 49.05 & 47.36 & 50.30 \\
\bottomrule
\end{tabular}
}
\caption{nDCG@10 on BEIR NQ.}
\label{tab:ablation_t_alpha}
\end{table}

Figure~\ref{fig:ablation_t_alpha} examines the temporal subspace dimension \(t\) and loss weighting coefficient \(\alpha\), \eq{eq:final_loss}, with auxiliary losses disabled; corresponding semantic performance on NQ is reported in Table~\ref{tab:ablation_t_alpha}. 
For \(\alpha\), values above 0.5 notably degrade \texttt{bge} performance, while \texttt{contriever} shows improved temporal retrieval up to \(\alpha = 0.5\) alongside a semantic trade-off. 
Regarding \(t\), \texttt{bge} requires \(t \geq 64\) to improve temporal retrieval without notable semantic loss. \texttt{contriever} follows a similar pattern, where increasing \(t\) yields consistent temporal gains with limited semantic impact. 
Based on these observations, we set \(\alpha \in [0.1, 0.25]\) and \(t \in [64, 128]\). 
Results on TimeQA are provided in Appendix~\ref{appendix_additional_hyperparam_ablation_studies_on_timeqa} and per-model configurations are detailed in Appendix~\ref{appendix:implementation_details}.

\paragraph{Auxiliary Losses Ablation Study.}
\label{par:auxiliary_loss_ablation}
\begin{table}[ht]
\setlength{\tabcolsep}{1.5pt}
\renewcommand{\arraystretch}{1.0}
\footnotesize
\centering
\scalebox{0.75}{
\begin{tabular}{cc|ccccc|ccccc|c}
\toprule
Dist. & CKA & \multicolumn{5}{c|}{nDCG{@}10} & \multicolumn{5}{c|}{Recall{@}100} & \multirow{2}{*}{NQ} \\
\cmidrule{3-7} \cmidrule{8-12}
\(\beta\) & \(\gamma\) & 64 & 128 & 256 & 512 & 768 & 64 & 128 & 256 & 512 & 768 \\
\midrule
0 & 0 & 36.5 & 51.3 & 62.7 & 66.0 & \textbf{68.8} & 87.9 & 95.8 & 97.7 & 97.9 & 98.3 & \textbf{19.6} \\
0.1 & 0 & 39.2 & \textbf{53.8} & 63.6 & 66.3 & 68.6 & 90.5 & 95.9 & 97.7 & 97.9 & 98.1 & 18.2 \\
0 & 0.1 & 37.3 & 52.3 & 63.5 & \textbf{66.4} & \textbf{68.8} & 88.6 & \textbf{96.0} & \textbf{97.8} & \textbf{98.0} & \textbf{98.4} & 19.4 \\
0.1 & 0.1 & \textbf{39.3} & \textbf{53.8} & \textbf{63.8} & 66.2 & 68.4 & 90.6 & \textbf{96.0} & 97.5 & 97.8 & 98.1 & 18.0 \\
0.25 & 0.25 & 37.4 & 48.7 & 55.9 & 58.4 & 60.4 & \textbf{91.0} & 95.0 & 96.9 & 97.2 & 97.4 & 14.2 \\
0.5 & 0.5 & 11.4 & 14.0 & 17.7 & 21.0 & 22.8 & 53.7 & 70.6 & 82.7 & 87.4 & 89.6 & 18.1 \\
\bottomrule
\end{tabular}
}
\caption{\texttt{contriever} auxiliary loss ablation on TNP.}
\label{tab:contriever_ablation_cka_distill}
\end{table}

\begin{table}[ht]
\setlength{\tabcolsep}{1.5pt}
\renewcommand{\arraystretch}{1.0}
\footnotesize
\centering
\scalebox{0.7}{
\begin{tabular}{cc|ccccc|ccccc|c}
\toprule
Dist. & CKA & \multicolumn{5}{c|}{nDCG{@}10} & \multicolumn{5}{c|}{Recall{@}100} & \multirow{2}{*}{NQ} \\
\cmidrule{3-7} \cmidrule{8-12}
\(\beta\) & \(\gamma\) & 64 & 128 & 256 & 512 & 768 & 64 & 128 & 256 & 512 & 768 \\
\midrule
0 & 0 & 47.2 & 63.6 & 72.9 & 76.6 & 77.3 & 88.4 & 94.6 & \textbf{96.4} & \textbf{97.1} & \textbf{97.0} & 50.3 \\
0.1 & 0 & 47.6 & 64.0 & 73.1 & 76.7 & 77.4 & 88.6 & 94.6 & \textbf{96.4} & \textbf{97.1} & \textbf{97.0} & \textbf{50.4} \\
0 & 0.1 & 47.7 & 64.0 & 73.3 & 76.8 & 77.6 & 88.8 & 94.6 & \textbf{96.4} & \textbf{97.1} & 96.9 & 49.9 \\
0.1 & 0.1 & \textbf{48.1} & \textbf{64.3} & \textbf{73.5} & \textbf{77.0} & \textbf{77.7} & \textbf{89.0} & \textbf{94.7} & \textbf{96.4} & \textbf{97.1} & 96.9 & 49.9 \\
0.25 & 0.25 & 31.7 & 47.0 & 54.6 & 58.4 & 59.4 & 65.9 & 80.8 & 87.2 & 89.7 & 90.6 & 49.6 \\
0.5 & 0.5 & 29.5 & 43.9 & 51.5 & 55.5 & 56.6 & 63.6 & 78.6 & 84.8 & 88.3 & 89.1 & 49.3 \\
\bottomrule
\end{tabular}
}
\caption{\texttt{bge} auxiliary loss ablation on TNP.}
\label{tab:bge_ablation_cka_distill}
\end{table}

Tables~\ref{tab:contriever_ablation_cka_distill} and~\ref{tab:bge_ablation_cka_distill} analyze the effect of the local (\(\beta\)) and structural (\(\gamma\)) self-distillation auxiliary losses in \eq{eq:final_loss} on the TNP dataset.
Setting both \(\beta\) and \(\gamma\) to a small value (e.g., 0.1) improves alignment at both local and global levels compared to using no auxiliary loss.
This setting is robust across \texttt{contriever} and \texttt{bge}: the former trades semantic for temporal retrieval, while the latter shows only minor changes.
However, larger values introduce competition with the contrastive objective.
We set \(\beta = \gamma = 0.1\) in all experiments.
Results on TimeQA are provided in Appendix~\ref{appendix_additional_hyperparam_ablation_studies_on_timeqa} with similar observations.

\section{Related Work}

\paragraph{Temporal Information Retrieval (TIR) \& Temporal Question Answering (TQA).}
TIR retrieves documents satisfying both semantic relevance and temporal constraints, while TQA extends this to time-aware reasoning~\cite{piryaniArxiv25ItsHigh2025}. Despite advances in dense retrieval~\cite{qwen3embedding,leeICLR25NVEmbedImproved2025} and TQA~\cite{fatemiICLR25TESTTIME2025,bazaga_reasonovertime_acl25}, integrating temporal signals into dense retrievers remains underexplored. Recent efforts include supervised temporal contrastive learning~\cite{wuCIKM24TimeSensitveRetrievalAugmented2024}, semantic–temporal fusion~\cite{abdallahArxiv25TempRetrieverFusionbasedTemporal2025}, and multi-retriever ensembles~\cite{han2025temporal}, alongside emerging temporal RAG frameworks~\cite{qianEMNLP24TimeR4Timeaware2024,zhang_mrag_emnlp25}. In contrast, TMRL is a single-model, plug-and-play approach that equips TEMs with temporal-aware Matryoshka embeddings, improving both TIR and Temporal RAG efficiency.

\paragraph{Matryoshka Representation Learning (MRL).}
\label{par:mrl}
Modern TEMs increasingly adopt MRL~\citep{kusupatiNeurIPS22MatryoshkaRepresentation} to produce nested embeddings for flexible compute-accuracy trade-offs~\cite{lee_gecko_arxiv24,nussbaumTMLR25NomicEmbed2025}. While M-Adaptor~\cite{yoonEMNLP24MatryoshkaAdaptorUnsupervised2024} and subsequent multi-stage variants~\cite{zhang_smec_emnlp25} adapt MRL for general retrieval efficiency.
TMRL shares a similar motivation as M-Adaptor: both use parameter-efficient finetuning to equip TEMs with Matryoshka embeddings. However, TMRL emphasizes TIR, whereas M-Adaptor preserves general semantic retrieval.
\section{Conclusion}
We propose Temporal-aware Matryoshka Representation Learning (TMRL), an efficient Matryoshka adaptation method that equips TEMs with temporal-aware embeddings, improving temporal retrieval while preserving semantic performance.
We also introduce a temporal contrastive data augmentation pipeline that provides cleaner temporal supervision.
Experiments show that TMRL enhances temporal encoding in retrievers, improves downstream temporal RAG performance, and enables flexible accuracy–efficiency trade-offs under varying computational constraints.

\section*{Limitations}
Despite evaluating TMRL across six widely used encoder-only text embedding models, we do not evaluate our method on LLM-based embedding models due to resource constraints. 
Evaluation on large-scale news, legal, or policy corpora, which often lack explicit temporal annotations, remains a promising direction for future work. 
However, it is worth noting that many real-world corpora are either access-restricted (e.g., copyrighted news) or lack time-sensitive retrieval annotations, making controlled evaluation of temporal retrieval difficult.
Finally, to ensure clean training signals, our augmented temporal contrastive learning dataset follows a simple principle: splitting long passages into shorter passages and discarding sentences containing multiple temporal expressions. 
Extending TMRL’s training augmentation strategy to explicitly learn from passages containing multiple temporal expressions, which is common in real-world documents, remains an open direction for future research.

\section*{Acknowledgments}
This material is based on research sponsored by Defense Advanced Research Projects Agency (DARPA) under agreement number HR0011-22-2-0047. The U.S. Government is authorized to reproduce and distribute reprints for Governmental purposes notwithstanding any copyright notation thereon. The views and conclusions contained herein are those of the authors and should not be interpreted as necessarily representing the official policies or endorsements, either expressed or implied, of DARPA or the U.S. Government.
This work is supported by Monash eResearch capabilities, including M3.
This work is partially supported by the Australian Research Council through Discovery Early Career Researcher Award (DE250100032). 
The authors are grateful to the anonymous reviewers for their helpful comments.

\bibliography{custom}
\appendix
\section{Examples for Figure~\ref{fig:1}}
\label{appendix:examples_fig1}

We use a question-answer pair from Temporal Nobel Prize (TNP) to demonstrate our argument in the introduction (Section~\ref{sec:introduction}).
\begin{quote}
    \textbf{Query}: Who won the Nobel Prize in Chemistry between 1910 and 1911? \\
    \textbf{Answer}: "Otto Wallach", "Marie Curie"
\end{quote}

Zero-shot \texttt{contriever} retrieves temporally misaligned documents (e.g., Rutherford 1908, Richards 1914), leading all generators to fail:
\begin{itemize}[nosep, leftmargin=8pt]
    \item \textbf{Qwen3-8B}: None (F1=0)
    \item \textbf{Qwen3-32B}: No one won the Nobel Prize ... (F1=0)
    \item \textbf{DeepSeek-R1-Distill-Qwen-32B}: None of the provided documents list... (F1=0)
    \item \textbf{Llama-3.3-70B-Instruct}: E. Rutherford. (F1=0)
\end{itemize}

With TMRL \texttt{contriever}, we correctly retrieved one relevant document from the top-5 retrieved documents:
\begin{enumerate}[nosep, leftmargin=12pt]
    \item \textbf{Otto Wallach (1910 Chemistry) (Correct)}
    \item Jacobus H. van't Hoff (1901 Chemistry)
    \item E. Rutherford (1908 Chemistry)
    \item W. Ostwald (1909 Chemistry)
    \item A. Kossel (1910 Medicine)
\end{enumerate}

As a result, stronger LLMs can leverage their strength to answer more correctly:
\begin{itemize}[nosep, leftmargin=8pt]
    \item \textbf{Qwen3-8B}: Otto Wallach and Jacobus H. van't Hoff (F1=0.33)
    \item \textbf{Qwen3-32B}: Otto Wallach, A. Kossel (F1=0.5)
    \item \textbf{DeepSeek-R1-Distill-Qwen-32B}: Otto Wallach (F1=0.67)
    \item \textbf{Llama-3.3-70B-Instruct}: Otto Wallach (F1=0.67)
\end{itemize}
\section{Additional Results}
\label{appendix:addition_results}

\subsection{TimeQA Main Results}
\label{appendix:timeqa_main_results}

\begin{figure}[ht]
    \centering
    \includegraphics[width=1.0\linewidth]{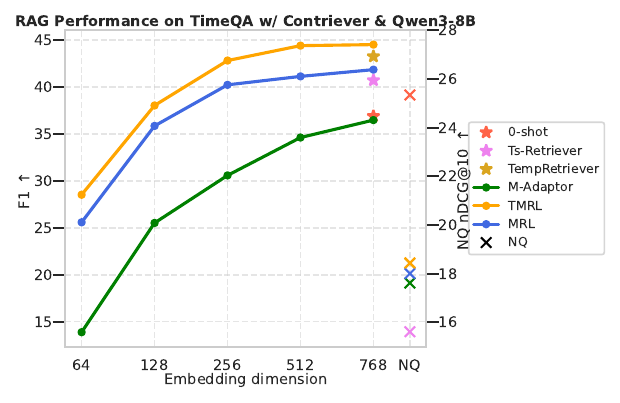}
    \caption{TimeQA RAG performance using Qwen3-8B with different \texttt{contriever} variants trained on our augmented data. Note that the respective semantic retrieval performance of TempRetriever and TMRL on NQ are 18.42 and 18.44, causing overlapping markers.}
    \label{fig:rag_timeqa}
\end{figure}

\begin{figure*}[!ht]
    \centering
    \setlength{\tabcolsep}{0.0pt}
    \renewcommand{\arraystretch}{0.0}
    \includegraphics[width=\linewidth]{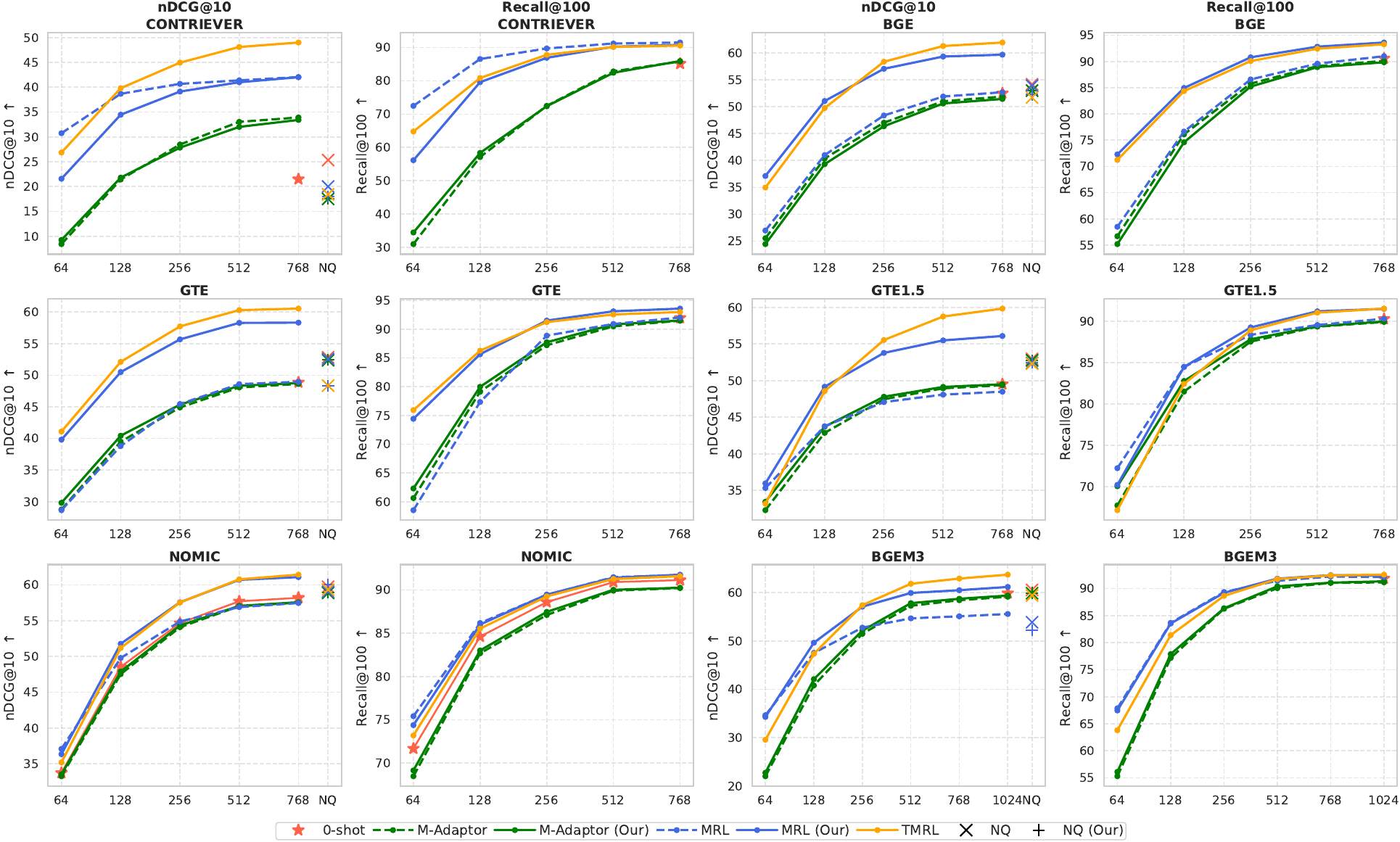}
    \caption{Matryoshka embedding results on TimeQA. (Our) denotes training using our augmented data.}
    \label{fig:matryoshka_result_timeqa}
\end{figure*}

Due to limited space, we attach the TimeQA results in this Appendix section.
The Matryoshka embedding results are displayed in Figure~\ref{fig:matryoshka_result_timeqa}.
The RAG result is presented in Figure~\ref{fig:rag_timeqa}.

\subsection{Additional Hyperparamaters Ablation Studies on TimeQA}
\label{appendix_additional_hyperparam_ablation_studies_on_timeqa}
\begin{table}[ht]
\setlength{\tabcolsep}{1.5pt}
\renewcommand{\arraystretch}{1.0}
\centering
\scalebox{0.8}{
\begin{tabular}{l|ccccc|ccccc|c}
\toprule
\multirow{2}{*}{\(\alpha\)} & \multicolumn{5}{c|}{nDCG{@}10} & \multicolumn{5}{c|}{Recall{@}100} & \multirow{2}{*}{NQ} \\
\cmidrule{2-6} \cmidrule{7-11}
& 64 & 128 & 256 & 512 & 768 & 64 & 128 & 256 & 512 & 768 \\
\midrule
0.1 & \textbf{64.8} & \textbf{80.8} & \textbf{87.8} & \textbf{90.2} & \textbf{90.6} & \textbf{26.9} & \textbf{39.8} & \textbf{45.0} & \textbf{48.1} & \textbf{49.0} & \textbf{18.4} \\
0.25 & 60.0 & 78.4 & 86.5 & 89.5 & 90.1 & 23.0 & 35.3 & 40.9 & 44.5 & 45.4 & \textbf{18.4} \\
0.5 & 57.6 & 76.1 & 84.8 & 88.0 & 88.9 & 21.3 & 33.1 & 39.3 & 42.9 & 44.2 & 18.2 \\
1 & 52.3 & 71.3 & 79.8 & 84.1 & 85.0 & 18.1 & 28.4 & 33.6 & 37.8 & 38.6 & 17.7 \\
\bottomrule
\end{tabular}
}
\caption{Ablation of the temporal subspace dimension \(t\) and the temporal subspace coefficient \(\alpha\) on TimeQA using \texttt{contriever}. Auxiliary losses are disabled (\(\beta = \gamma = 0\)).}
\label{tab:ablation_t_alpha_contriever_timeqa}
\end{table}

\begin{table}[ht]
\setlength{\tabcolsep}{1.5pt}
\renewcommand{\arraystretch}{1.0}
\centering
\scalebox{0.8}{
\begin{tabular}{l|ccccc|ccccc|c}
\toprule
\multirow{2}{*}{\(\alpha\)} & \multicolumn{5}{c|}{nDCG{@}10} & \multicolumn{5}{c|}{Recall{@}100} & \multirow{2}{*}{NQ} \\
\cmidrule{2-6} \cmidrule{7-11}
& 64 & 128 & 256 & 512 & 768 & 64 & 128 & 256 & 512 & 768 \\
\midrule
0.1 & 70.5 & 84.2 & \textbf{90.5} & \textbf{92.4} & \textbf{93.5} & 34.6 & \textbf{49.9} & 57.8 & 60.6 & 61.1 & \textbf{52.3} \\
0.25 & \textbf{71.2} & \textbf{84.4} & 90.1 & \textbf{92.4} & 93.3 & \textbf{35.0} & 49.7 & \textbf{58.3} & \textbf{61.3} & \textbf{61.9} & 51.7 \\
0.5 & 62.3 & 78.1 & 87.4 & 90.5 & 91.6 & 25.9 & 40.7 & 51.3 & 55.4 & 55.7 & 51.8 \\
\bottomrule
\end{tabular}
}
\caption{Ablation of the temporal subspace dimension \(t\) and the temporal subspace coefficient \(\alpha\) on TimeQA using \texttt{bge}. Auxiliary losses are disabled (\(\beta = \gamma = 0\)).}
\label{tab:ablation_t_alpha_bge_timeqa}
\end{table}

\begin{table}[ht]
\setlength{\tabcolsep}{1.5pt}
\renewcommand{\arraystretch}{1.0}
\centering
\scalebox{0.72}{
\begin{tabular}{cc|ccccc|ccccc|c}
\toprule
Dist. & CKA & \multicolumn{5}{c|}{nDCG{@}10} & \multicolumn{5}{c|}{Recall{@}100} & \multirow{2}{*}{NQ} \\
\cmidrule{3-7} \cmidrule{8-12}
\(\beta\) & \(\gamma\) & 64 & 128 & 256 & 512 & 768 & 64 & 128 & 256 & 512 & 768 \\
\midrule
0 & 0 & 55.3 & 78.9 & 87.0 & \textbf{90.2} & \textbf{90.8} & 20.8 & 34.0 & 39.1 & 41.4 & 42.2 & 18.0 \\
0.1 & 0.1 & \textbf{64.5} & \textbf{80.8} & \textbf{87.7} & 90.1 & 90.5 & \textbf{26.7} & \textbf{39.6} & \textbf{44.7} & \textbf{47.9} & \textbf{48.8} & \textbf{18.4} \\
0.25 & 0.25 & 49.7 & 67.1 & 79.1 & 85.9 & 87.0 & 15.9 & 25.5 & 30.0 & 34.1 & 35.1 & 18.1 \\
0.5 & 0.5 & 33.7 & 56.0 & 70.1 & 81.9 & 84.7 & 9.3 & 19.8 & 24.6 & 30.3 & 31.6 & 18.0 \\
\bottomrule
\end{tabular}
}
\caption{\texttt{contriever} auxiliary loss ablation on TimeQA.}
\label{tab:contriever_ablation_cka_distill_timeqa}
\end{table}

\begin{table}[ht]
\setlength{\tabcolsep}{1.5pt}
\renewcommand{\arraystretch}{1.0}
\centering
\scalebox{0.72}{
\begin{tabular}{cc|ccccc|ccccc|c}
\toprule
Dist. & CKA & \multicolumn{5}{c|}{nDCG{@}10} & \multicolumn{5}{c|}{Recall{@}100} & \multirow{2}{*}{NQ} \\
\cmidrule{3-7} \cmidrule{8-12}
\(\beta\) & \(\gamma\) & 64 & 128 & 256 & 512 & 768 & 64 & 128 & 256 & 512 & 768 \\
\midrule
0 & 0 & 62.4 & 79.0 & 88.2 & 91.1 & 91.9 & 27.2 & 42.7 & 52.2 & 55.7 & 56.0 & \textbf{52.3} \\
0.1 & 0.1 & \textbf{70.5} & \textbf{84.2} & \textbf{90.5} & \textbf{92.4} & \textbf{93.5} & \textbf{34.6} & \textbf{49.9} & \textbf{57.8} & \textbf{60.6} & \textbf{61.1} & \textbf{52.3} \\
0.25 & 0.25 & 61.8 & 78.1 & 87.7 & 90.6 & 91.5 & 26.2 & 41.3 & 51.0 & 54.9 & 55.1 & 52.2 \\
0.5 & 0.5 & 65.5 & 80.7 & 88.6 & 91.3 & 92.4
 & 29.1 & 43.9 & 53.4 & 57.3 & 57.6 & 52.2 \\
\bottomrule
\end{tabular}
}
\caption{\texttt{bge} auxiliary loss ablation on TimeQA.}
\label{tab:bge_ablation_cka_distill_timeqa}
\end{table}

We extend our hyperparamaters ablation studies to the TimeQA dataset, which represents a fundamentally different domain distribution: 
while TNP focuses on isolated, fact-based temporal retrieval with explicit dates, TimeQA involves complex, evolving entity histories from Wikipedia where temporal signals are deeply entangled with dense semantic context.

\paragraph{Ablation studies on \(\alpha\).}
According to the results from Table~\ref{tab:ablation_t_alpha_contriever_timeqa} and~\ref{tab:ablation_t_alpha_bge_timeqa}, we conclude that the best  for \texttt{contriever} and \texttt{bge} on TimeQA are 0.1 and 0.25, respectively. 

\paragraph{Ablation studies on \(\beta, \gamma\).}
According to the results from  Table~\ref{tab:contriever_ablation_cka_distill_timeqa} and~\ref{tab:bge_ablation_cka_distill_timeqa}, we conclude that \(\beta = \gamma = 0.1\) provides the best performance for both \texttt{contriever} and \texttt{bge} on TimeQA.

The above ablation studies demonstrate that default values \(\alpha \in [0.1, 0.25]\) and \(\beta = \gamma = 0.1\) in the main paper are robust enough for both \texttt{contriever} and \texttt{bge} on a different domain distribution - TimeQA, where they consistently perform better compared to other settings.

\subsection{Ablation Study on Allen Relations}
\label{appendix_ablation_study_allen_relations}
\begin{table*}[ht]
\centering
\scalebox{0.9}{
\begin{tabular}{l|*{6}{S[table-format=-2.1]}S[table-format=4.0]}
\toprule
Data variants & {64} & {128} & {256} & {512} & {768} & {BEIR NQ} & {\# training samples} \\
\midrule
\textbf{All (TMRL \texttt{contriever})} & 26.9 & 39.8 & 45.0 & 48.1 & 49.0 & 18.4 & 8149 \\
\midrule
No Before/After & -15.2 & -16.6 & -15.5 & -14.1 & -13.6 & -0.4 & 1849 \\
No Equals & -0.5 & -1.4 & -1.3 & -2.0 & -2.1 & 0.6 & 8146 \\
No Contains/During & -6.7 & -6.1 & -5.5 & -4.8 & -4.5 & 0.8 & 8124 \\
No Overlaps/Overlapped-by & -7.4 & -7.9 & -7.7 & -8.3 & -8.1 & -0.2 & 8149 \\
No Finishes/Finished-by & 0.5 & 0.6 & 0.5 & 0.4 & 0.5 & 0.1 & 8149 \\
No Meets/Met-by & 0.3 & 0.3 & 0.3 & 0.3 & 0.3 & 0.0 & 8149 \\
No Starts/Started-by & 0.3 & 0.5 & 0.4 & 0.4 & 0.6 & 0.2 & 8149 \\
No ``Temporal Answer'' Queries & -4.9 & -5.1 & -5.1 & -5.7 & -5.7 & -0.5 & 8114 \\
\bottomrule
\end{tabular}
}
\caption{Ablation study on TimeQA and BEIR NQ using TMRL \texttt{contriever}. Values report difference in nDCG{@}10 relative to the model train with the fully augmented data (All) (positive/negative values indicate performance gain/drop). The final column shows the number of training samples containing at least one positive/negative query after filtering.}
\label{tab:allen_relation_ablation}
\end{table*}

To analyze how TMRL captures temporal reasoning patterns, we systematically remove each category of Allen relations from the augmented TimeQA dataset (both positives and negatives queries) and report the difference in retrieval performance on TimeQA, which contains diverse temporal query types. This controlled removal allows us to quantify how different temporal Allen relations contribute to the learned representations. We group symmetrical Allen relations together, according to~\citet{islakoglu-kalo-2025-chronosense}.

From Table~\ref{tab:allen_relation_ablation}, we observe that Before/After relations dominate the training data and are critical, removing them leads to the largest performance drop. Relations involving exact timestamps (Equals) and temporal intervals (Contains/During, Overlaps/Overlapped-by) also substantially improve temporal retrieval, though with a mild trade-off against purely semantic retrieval. In contrast, Finishes/Finished-by, Meets/Met-by, and Starts/Started-by are rarely generated by our pipeline, and their removal does not harm (and slightly improves) performance on TimeQA.

Additionally, our augmentation includes “Temporal Answer” queries that ask for event timestamps without explicit temporal constraints, which further strengthen temporal representations while preserving semantic retrieval.

\subsection{\texttt{Qwen3-Embedding-0.6B}'s Preliminary Results on TNP}
\label{appendix:qwen3_embedding_0.6b_results}
We extended our evaluation on the TNP dataset to \texttt{Qwen3-Embedding-0.6B}~\cite{qwen3embedding}, a representative instruction-tuned LLM-based embedder, as preliminary results beyond the encoder-only TEMs.
Table~\ref{tab:qwen3_embedding_0.6b_results} shows that there is a divergence in representation learning dynamics when applied TMRL to instruction-tuned architectures: 
setting the auxiliary structural loss coefficients to zero (\(\beta=\gamma=0\)) consistently outperforms the standard setting (\(\beta=\gamma=0.1\)) used for encoder-only TEMs.
We hypothesize that while encoder-only bi-encoders (e.g., \texttt{contriever}) require structural mimicry via CKA and self-distillation to align truncated dimensions (as shown in Tables~\ref{tab:contriever_ablation_cka_distill} and~\ref{tab:bge_ablation_cka_distill}), instruction-tuned LLM-based embedders possess a highly optimized, instruction-aligned representational geometry. Forcing these models to structurally mimic the full-dimensional space causes over-regularization.

The funnel retrieval approach remains applicable to \texttt{Qwen3-Embedding-0.6B}. As shown in Table~\ref{tab:qwen3_0.6b_embedding_funnel_retrieval}, it maintains comparable nDCG{@}10 performance while reducing latency compared to standard full-dimensional retrieval.

\begin{table*}[ht]
\renewcommand{\arraystretch}{1.0}
\centering
\begin{tabular}{l|ccccccc}
\toprule
Method & 64 & 128 & 256 & 512 & 768 & 1024 & NQ \\
\midrule
\multicolumn{7}{l}{\textbf{Recall{@}100}} \\
Zero-shot & 78.0 & 86.2 & 91.6 & 94.1 & 94.9 & 95.3 & - \\
MRL & 84.0 & 89.6 & \textbf{93.2} & \textbf{95.2} & \textbf{95.8} & \textbf{95.7} & - \\
TMRL (\(\beta=\gamma=0.1\)) & 84.6 & 89.7 & 92.6 & 94.6 & 95.1 & 95.3 & - \\
TMRL (\(\beta=\gamma=0\)) & \textbf{86.2} & \textbf{90.6} & 93.1 & 94.8 & 95.3 & 95.5 & - \\
\midrule
\multicolumn{6}{l}{\textbf{nDCG{@}10}} \\
Zero-shot & 17.8 & 27.0 & 36.8 & 44.7 & 49.9 & 52.4 & \textbf{53.5} \\
MRL & 39.3 & 52.1 & 61.8 & 66.4 & 67.3 & \textbf{67.3} & 39.5 \\
TMRL (\(\beta=\gamma=0.1\)) & 42.0 & 53.5 & 61.4 & 65.6 & 66.2 & 66.1 & 41.8 \\
TMRL (\(\beta=\gamma=0\)) & \textbf{44.3} & \textbf{54.8} & \textbf{62.8} & \textbf{66.9} & \textbf{67.5} & \textbf{67.3} & 39.6 \\
\bottomrule
\end{tabular}
\caption{\texttt{Qwen3-Embedding-0.6B} results on TNP.}
\label{tab:qwen3_embedding_0.6b_results}
\end{table*}

\begin{table*}[ht]
\renewcommand{\arraystretch}{1.0}
\centering
\begin{tabular}{l|ccc}
\toprule
Method & nDCG{@}10 & Latency (s) & Latency on 10\(\times\) larger corpus (s) \\
\midrule
TMRL & 67.34 & 0.090 & 0.154 \\
Funnel TMRL & \textbf{67.35} & \textbf{0.086} & \textbf{0.105} \\
\bottomrule
\end{tabular}
\caption{Funnel retrieval (top-k=100, d=64) \(\rightarrow\) (top-k=10, d=1024) vs. standard full-dimensional retrieval using \texttt{Qwen3-Embedding-0.6B}.}
\label{tab:qwen3_0.6b_embedding_funnel_retrieval}
\end{table*}

\subsection{Semantic Retrieval Results on BEIR Datasets}
Following TsRetriever~\cite{wuCIKM24TimeSensitveRetrievalAugmented2024} and TSM~\cite{han2025temporal}, we use Natural Questions (NQ) as a benchmark to evaluate semantic retrieval preservation. NQ contains 3452 test queries over 2.68M passages, of which only 1.54\% are time-sensitive~\cite{han2025temporal}. Another widely used benchmark is MS MARCO (7k test queries, 8.8M documents). However, as reported in TSM, TEMs that perform well on NQ generally exhibit similar trends on MS MARCO.
Nevertheless, we additionally report results on additional datasets from the BEIR benchmark~\cite{thakurBEIRHeterogeneousBenchmark}, including SciFact and MS MARCO in terms of nDCG@10, using \texttt{contriever} (Table~\ref{tab:contriever_semantic}) and \texttt{bge} (Table~\ref{tab:bge_semantic}). The results show that TMRL preserves semantic retrieval performance across datasets from the BEIR benchmark, with behavior varying slightly depending on the underlying TEM.
\begin{table*}[ht]
\setlength{\tabcolsep}{1.0pt}
\renewcommand{\arraystretch}{1.0}
\centering
\scalebox{1.0}{
\begin{tabular}{l|c|c|c}
\toprule
Dataset & \texttt{contriever} (zero-shot) & TMRL \texttt{contriever} & Semantic Preservation Ratio (\%) $\uparrow$ \\
\midrule 
NQ & 25.4 & 18.0 & 71\% \\
SciFact & 57.1 & 53.2 & 93\% \\
MS MARCO & 20.6 & 15.9 & 77\% \\
\bottomrule
\end{tabular}
}
\caption{Semantic retrieval preservation in terms of nDCG@10 between zero-shot \texttt{contriever} and TMRL \texttt{contriever} on NQ, SciFact, and MS MARCO from BEIR.}
\label{tab:contriever_semantic}
\end{table*}

\begin{table*}[ht]
\setlength{\tabcolsep}{1.0pt}
\renewcommand{\arraystretch}{1.0}
\centering
\scalebox{1.0}{
\begin{tabular}{l|c|c|c}
\toprule
Dataset & \texttt{bge} (zero-shot) & TMRL \texttt{bge} & Semantic Preservation Ratio (\%) $\uparrow$ \\
\midrule 
NQ & 54.2 & 50.0 & 92\% \\
SciFact & 74.3 & 72.0 & 97\% \\
MS MARCO & 35.0 & 33.8 & 97\% \\
\bottomrule
\end{tabular}
}
\caption{Semantic retrieval preservation in terms of nDCG@10 between zero-shot \texttt{bge} and TMRL \texttt{bge} on NQ, SciFact, and MS MARCO from the BEIR benchmarks.}
\label{tab:bge_semantic}
\end{table*}

\subsection{Qualitative Failure Analysis}
We conducted manual case studies on TimeQA to identify specific failure modes of TMRL compared to baselines. 
Our analysis reveals that TMRL faces greater challenges in handling complex cases where temporal constraints are deeply intertwined with semantic inference. 
While baselines also struggle with these cases, they achieve higher Recall{@}k (for larger k) because they tend to retrieve a broader range of semantically related context. 
In contrast, TMRL excels at strictly ranking the most temporally relevant passages in the top positions, directly driving its superior nDCG{@}10 performance. Here are two representative cases:

\paragraph{Case 1: TMRL fails, MRL succeeds.}
\begin{itemize}[nosep,leftmargin=8pt]
    \item \textbf{Query}: ``Which political party did Walter Veltroni belong to between Jan 1999 and Aug 2006?''
    \item \textbf{Gold Passage}: ``In 1998, he resigned, subsequent to his election as National Secretary of the Democrats of the Left (DS)... For example, in 2001, after the late night show ``Satyricon'' aired an interview that discussed indictments on links between the right-wing leader and the mafia, Marco Travaglio reported that Veltroni dispatched a messenger menacing the closure of the show.''
    \item \textbf{Analysis}: The gold passage lacks an explicit temporal expression matching ``1999–2006''. Retrieving it requires inferring that his role elected in 1998 continued through 2006, based on the given semantic context and corresponding temporal expression “in 2001”.
    \item \textbf{MRL}: Ranks the gold passage 10th. which indicates that MRL tried to retrieve everything related to politics (i.e., the semantic context).
    \item \textbf{TMRL}: Fails to retrieve it in the Top-10.
\end{itemize}

\paragraph{Case 2: TMRL succeeds, MRL fails}

\begin{itemize}[leftmargin=8pt]
    \item \textbf{Query}: ``Which team did the player Aman Verma belong to between Jul 2009 and Sep 2010?''
    \item \textbf{Gold Passage}: ``Section: Leicester City... Verma had his first call up to the main Leicester squad as an unused substitute in a 2–1 win at Blackpool on 6 February 2010. He joined Conference National club Histon on loan for the remainder of the 2009–10 season on 26 March 2010, In May 2010, Verma was awarded another contract extension by Leicester.''
    \item \textbf{Analysis}: The corpus contains over ten highly semantically similar passages about Aman Verma's career across different years.
    \item \textbf{MRL}: Fails to retrieve the gold passage in the Top-10. It gets distracted by semantically identical but temporally mismatched passages (e.g., Top-1 discusses events in 2011 while the Top-2 provides an abstract overview of the player).
    \item \textbf{TMRL}: Successfully ranks the gold passage at Top-5. It correctly prioritizes the passage containing the 2010 temporal signals.
\end{itemize}
\section{Proof of Matryoshka Representation Learning as Multi-Scale Mutual Information Maximization} 
\label{appendix:proof}
Consider the setup in Section~\ref{subsubsec:temporal_mrl}, where the input is a passage representation \(\vect{z}_p\) and the target is a query representation \(\vect{z}_q\).
Matryoshka Representation Learning (MRL) aggregates the InfoNCE loss across all target dimensions:
\begin{equation}
    \mathcal{L}_{\mathrm{MRL}} = \sum_{m \in \mathcal{M}} \mathcal{L}_{\mathrm{InfoNCE}}^{(m)}.
\end{equation}
Given a positive pair \((\vect{z}_p, \vect{z}_{q^+}) \sim p_{\mathrm{pos}}(\vect{z}_p, \vect{z}_q)\) and \(K-1\) negative samples \(\{\vect{z}_{q_i^-}\}_{i=1}^{K-1} \sim p(\vect{z}_q)\) drawn independently from the marginal distribution of the queries, the InfoNCE loss at scale \(m\) is:
\begin{equation}
\begin{gathered}
    \mathcal{L}_{\mathrm{InfoNCE}}^{(m)} \triangleq -\mathbb{E} \\
    \left[ \log \frac{e^{\mathrm{sim}^{(m)}(\vect{z}_p, \vect{z}_{q^+}) / \tau}}{e^{\mathrm{sim}^{(m)}(\vect{z}_p, \vect{z}_{q^+}) / \tau} + \sum\limits_{i=1}^{K-1} e^{\mathrm{sim}^{(m)}(\vect{z}_p, \vect{z}_{q_i^-}) / \tau}} \right].
\end{gathered}
\end{equation}
Following the variational mutual information bound established in contrastive representation learning~\citep{oord_infonce_arxiv18, poole_mutualinformationbound_icml19}, minimizing $\mathcal{L}_{\mathrm{InfoNCE}}^{(m)}$ maximizes a lower bound on the mutual information preserved in the truncated representation:
\begin{equation}
    I\left((\vect{z}_p)_{\leq m}; (\vect{z}_q)_{\leq m}\right) \geq \log K - \mathcal{L}_{\mathrm{InfoNCE}}^{(m)}.
\end{equation}
Summing this inequality over all scales $m \in \mathcal{M}$ yields:
\begin{align}
    \sum_{m \in \mathcal{M}} I\left((\vect{z}_p)_{\leq m}; (\vect{z}_q)_{\leq m}\right) \geq |\mathcal{M}| \log K - \mathcal{L}_{\mathrm{MRL}}.
\end{align}

Defining the tractable variational lower bound on cumulative mutual information as:
\begin{equation}
    \mathcal{I}_{\mathrm{MRL}} \triangleq |\mathcal{M}| \log K - \mathcal{L}_{\mathrm{MRL}},
\end{equation}
we have $\sum_{m \in \mathcal{M}} I\big((\vect{z}_p)_{\leq m}; (\vect{z}_q)_{\leq m}\big) \geq \mathcal{I}_{\mathrm{MRL}}$.
Since $|\mathcal{M}| \log K$ is constant with respect to the encoder parameters $\theta$, minimizing the MRL objective is equivalent to maximizing this lower bound:
\begin{equation}
\begin{gathered}
    \min_{\theta} \mathcal{L}_{\mathrm{MRL}} \ \Longleftrightarrow \ \max_{\theta}\, \mathcal{I}_{\mathrm{MRL}} \ \leq \\
    \max_{\theta} \sum_{m \in \mathcal{M}} I\left((\vect{z}_p)_{\leq m}; (\vect{z}_q)_{\leq m}\right).
\end{gathered}
\end{equation}
\section{Implementation Details}
\label{appendix:implementation_details}

\begin{table*}[ht]
\setlength{\tabcolsep}{1.0pt}
\renewcommand{\arraystretch}{1.0}
\centering
\scalebox{0.9}{
\begin{tabular}{l|C{3cm}|C{3cm}|C{3cm}|c|c}
\toprule
\textbf{TEM} & \textbf{HuggingFace Signature} & \textbf{Query's prefix} & \textbf{Passage's prefix} & \textbf{Temperature} \(\tau\) & \textbf{Pooling} \\
\midrule 
\texttt{contriever}          & {facebook/contriever} & & & 0.05 &  Mean \\
\midrule
\texttt{gte}                 & {thenlper/gte-base} & & & 0.02 &  Mean \\
\midrule
\texttt{nomic}               & {nomic-ai/nomic-embed-text-v1.5} & search\_query: & search\_document: & 0.05 &  Mean \\
\midrule
\texttt{bge}                 & {BAAI/bge-base-en-v1.5} & Represent this sentence for searching relevant passages: & Represent this sentence for searching relevant passages: & 0.02 &  CLS \\
\midrule
\texttt{gte1.5}              & {Alibaba-NLP/gte-base-en-v1.5} & & & 0.02 &  CLS \\
\midrule
\texttt{bgem3}               & {BAAI/bge-m3} & & & 0.02 &  CLS \\
\midrule
\texttt{qwen3} & {Qwen/Qwen3-Embedding-0.6B} & Instruct: Given a web search query, retrieve relevant passages that answer the query\textbackslash n Query: & & 0.02 & EOS \\
\bottomrule
\end{tabular}
}
\caption{Training hyperparameters of TEMs.}
\label{tab:tems_hyperparameters}
\end{table*}
\begin{table*}[ht]
\centering
\scalebox{1.0}{
\begin{tabular}{l|cccccccc}
\toprule
\multirow{2}{*}{\textbf{TEM}} & \multicolumn{4}{c}{\textbf{TNP}}    & \multicolumn{4}{c}{\textbf{TimeQA}} \\
\cmidrule(lr){2-5} \cmidrule(lr){6-9}
\multicolumn{1}{c|}{} & \(t\)   & $\alpha$ & $\beta$ & $\gamma$ & \(t\)   & $\alpha$ & $\beta$ & $\gamma$ \\
\midrule
\texttt{contriever}          & 64  & 0.1   & 0.1  & 0.1   & 128 & 0.1   & 0.1  & 0.1   \\
\texttt{gte}                 & 64  & 0.1   & 0.1  & 0.1   & 64  & 0.1   & 0.1  & 0.1   \\
\texttt{nomic}               & 128 & 0.1   & 0.1  & 0.1   & 128 & 0.1   & 0.1  & 0.1   \\
\texttt{bge}                 & 128 & 0.1   & 0.1  & 0.1   & 128 & 0.25  & 0.1  & 0.1   \\
\texttt{gte1.5}              & 64  & 0.25  & 0.1  & 0.1   & 128 & 0.1   & 0.1  & 0.1   \\
\texttt{bgem3}               & 128 & 0.1   & 0.1  & 0.1   & 128 & 0.1   & 0.1  & 0.1  \\
\texttt{qwen3}               & 128 & 0.1   & 0  & 0   & - & -   & -  & -  \\
\bottomrule
\end{tabular}
}
\caption{TMRL's training hyperparameters for each TEM.}
\label{tab:tmrl_hyperparameters}
\end{table*}

All experiments are conducted on a single NVIDIA A100 80GB GPU using the AdamW optimizer~\cite{loshchilov2019decoupled}.
All methods are fine-tuned for 5 epochs with in-batch negatives and additional hard negatives, except for \texttt{nomic}, which is only fine-tuned for 1 epoch, as noted in the original paper~\cite{nussbaumTMLR25NomicEmbed2025}.
We summarize the general configurations for the six text embedding models (TEMs) used for evaluation in Table~\ref{tab:tems_hyperparameters}.
We develop our methods and run experiments using the Tevatron framework~\cite{gao_tevatron_arxiv22}.
We use Pyserini~\cite{Lin_etal_SIGIR2021_Pyserini} to perform retrieval evaluation.
We use the vLLM framework~\cite{kwon_vllm_sigop23} for all LLMs related generation.

\subsection{BM25 and full fine-tuning baselines}
\begin{itemize}
    \item \textbf{BM25}: We use the default implementation of BM25s~\cite{lv_bm25s_arxiv24}.
    \item \textbf{Ts-Retriever}: We follow the original implementation.
    \item \textbf{TempRetriever}: TempRetriever employs two separate TEMs for semantic and temporal encoding and requires temporal expression extraction at inference time. We use SUTime~\cite{chang_sutime_2012} to extract all temporal expressions, which are concatenated using comma separation. As a result, TempRetriever has a larger model footprint and significantly higher inference latency than other methods. 
    As the code is not publicly available at the time of writing, we implement the Element-Wise Interaction (EWI) variant instead of Feature Stacking (FS) to ensure that all models using the same TEM have the same embedding dimensionality with the training setting reported in their paper.
\end{itemize}
For fair comparison, both Ts-Retriever and TempRetriever are evaluated without query routers, as our method does not employ one.

\subsection{Parameter-efficient fine-tuning baselines}
\begin{itemize}
    \item \textbf{LoRA}: We apply LoRA to all linear layers of the TEM. The LoRA rank is set to \(r=4\), the scaling factor is 4, and the dropout rate is 0.1.
    \item \textbf{Matryoshka-Adaptor}: As the code is not publicly available, we follow the original paper and a prior work from the same group~\cite{yoon_searchadaptor_acl24}. The adaptor is implemented as a linear layer with the same hidden dimension as the embedding and includes a skip connection. The adaptor is applied before the pooling operation.
    \item \textbf{MRL}: This method follows the same setup as \textbf{LoRA} and train with \(\mathcal{L}_{\mathrm{MRL}}\).
\end{itemize}

\subsection{Temporal-aware Matryoshka Representation Learning (TMRL)}
TMRL follows the same settings as the \textbf{MRL} baseline.
The projector \(\mathcal{P}\) is discarded after training.
We use \(k=10\) for \eq{eq:loss_pairwise_topk}.
The hyperparameters in \eq{eq:final_loss} are summarized in Table~\ref{tab:tmrl_hyperparameters}.

Regarding benchmarking retrieval latency, we take the average time of 100 retrieval process using precomputed query and corpus indices. Each iteration of retrieval only considers the approximate nearest neighbor search operation using FAISS~\cite{johnson_faiss_tbd19}.

\subsection{LLMs' settings for Temporal RAG}
\label{appendix:llm_settings_for_temporal_rag}
We use the FlashRAG framework~\cite{jinWWW25FlashRAGModular2025} and mainly \texttt{Qwen3-8B}~\cite{qwen3} as the LLM for generation to conduct RAG experiments.
We use the top-5 retrieved documents from the retriever as context input to the LLM.

\texttt{Qwen3-8B}'s and \texttt{Qwen3-32B}'s hyperparameters are as follows:
max\_tokens=8192,
temperature=0.7,
top\_p=0.8,
top\_k=20,
min\_p=0.

\texttt{DeepSeek-R1-Distill-Qwen-32B}'s hyperparameters are as follows:
max\_tokens=8192,
temperature=0.6,
top\_p=0.95,
top\_k=20,
min\_p=0.

\texttt{Llama-3.3-70B-Instruct-bnb-4bit}'s hyperparameters are as follows:
max\_tokens=8192,
temperature=0.6,
top\_p=0.9,
top\_k=20,
min\_p=0.

The system prompt is:
``Answer the question based on the given document. Only give me the answer and do not output any other words. The following are given documents. \{reference\}.''  
The user prompt is:
``Question: \{question\} Answer:''

\subsection{Details about temporal contrastive learning data augmentation}
\label{appendix:augmentation_pipeline}

We use \texttt{Qwen/Qwen3-4B-Thinking-2507}~\cite{qwen3} for generating the augmented positive/negative queries, instructed with a prompt template (Table~\ref{tab:prompt_template1}).
Within this prompt, we use the JSON schema and Allen relations defined in Tables~\ref{tab:json_schema} and~\ref{tab:allen_relation}, respectively.
The prompt includes few-shot examples drawn from the original TNP dataset, with temporal expressions extracted using SUTime.
We also provide the prompt used for constructing the TimeQA qrel files in Table~\ref{tab:timeqa_qrel_prompt}.

The LLM also extracts temporal spans from queries and passages, refined with preliminary SUTime outputs, without introducing new facts to preserve dataset fidelity.
The LLM hyperparameters are as follows:
max\_tokens=32768,
temperature=0.7,
top\_p=0.8,
top\_k=20,
min\_p=0.

\paragraph{Post-processing filtering analysis.}
\begin{table*}[ht]
\centering
\scalebox{0.9}{
\begin{tabular}{c|cccccc}
\toprule
Embedding dimension & Recall{@}100 (No filtering) \(\uparrow\) & Recall{@}100 (Filtered) \(\uparrow\) & Relative Gain (\%) \(\uparrow\) \\
\midrule
64 & 48.1 & 48.5 & 0.8 \\
128 & 65.0 & 65.7 & 1.1 \\
256 & 75.5 & 76.3 & 1 \\
512 & 77.9 & 78.2 & 0.4 \\
768 & 80.2 & 80.5 & 0.3 \\
\bottomrule
\end{tabular}
}
\caption{Impact of data filtering on Recall{@}100 performance of TMRL Contriever on TNP}
\label{tab:filtering_analysis_recall_100}
\end{table*}

\begin{table*}[ht]
\small
\centering
\scalebox{1.0}{
\begin{tabular}{c|cccccc}
\toprule
Embedding dimension & nDCG{@}10 (No filtering) \(\uparrow\) & nDCG{@}10 (Filtered) \(\uparrow\) & Relative Gain (\%) \(\uparrow\) \\
\midrule
64 & 38.8 & 39.3 & 1.2 \\
128 & 53.2 & 53.8 & 1.1 \\
256 & 63.1 & 63.8 & 1.1 \\
512 & 65.9 & 66.2 & 0.5 \\
768 & 68.0 & 68.4 & 0.5 \\
\bottomrule
\end{tabular}
}
\caption{Impact of data filtering on nDCG@10 performance of TMRL Contriever on TNP}
\label{tab:filtering_analysis_ndcg_10}
\end{table*}

After generation,  we apply a heuristic post-processing filtering step motivated by careful inspection of the raw outputs. 
These filtering steps are designed to reduce temporally irrelevant or ambiguous supervision and prevent bias toward the LLM reference year.
Specifically, we:
\begin{itemize}[nosep]
    \item Remove training samples with insufficient number of positive or negative query.
    \item Remove queries containing the current reference year used by the LLM (e.g., 2026), which may introduce artifacts in relative temporal expressions such as ``this year'', ``last year'', etc.
    \item Remove passages and queries containing ambiguous relative expressions such as ``now'', ``today'', ``yesterday'', etc.
\end{itemize}

In Tables~\ref{tab:filtering_analysis_recall_100} and~\ref{tab:filtering_analysis_ndcg_10}, we empirically verify the proposed post-processing filtering. After removing relative expressions and reference-year artifacts, TMRL \texttt{contriever} consistently achieves improved performance across embedding dimensions, supporting the effectiveness of the filtering process and indicating that any remaining noise does not dominate the training signal.
\section{Datasets Statistics}
\label{appendix:dataset_stats}

\begin{table*}[ht]
\centering
\scalebox{0.9}{
\begin{tabular}{lccccc}
\toprule

\multirow{2}{*}{\textbf{Temporal Nobel Prize}} &
\multicolumn{2}{c}{\textbf{Train}} &
\multicolumn{2}{c}{\textbf{Dev}} &
\multirow{2}{*}{\textbf{Test}} \\
\cmidrule(lr){2-3} \cmidrule(lr){4-5}
& Original & Our & Original & Our \\
\midrule
\# Unique documents & 8060 & 5922 & 165 & 306 & 989 \\
\# Passages & 8060 & 11693 & 165 & 1630 & 989 \\
\# Positive queries & 23388 & 64564 & 468 & 1520 & 3244 \\
\# Negative queries & 23388 & 60563 & 468 & 146 & - \\
\# Qrel & - & - & - & - & 13110 \\
   
\bottomrule
\end{tabular}
}
\caption{
    Dataset statistics of the original Temporal Nobel Prize dataset and our augmented version.
}
\label{tab:tnp_passage_query_count}
\end{table*}       

\begin{table*}[ht]
\centering
\scalebox{0.9}{
\begin{tabular}{l|cccccc}
\toprule
\textbf{Temporal Nobel Prize} & \multicolumn{3}{c}{\textbf{Train}} & \multicolumn{3}{c}{\textbf{Dev}} \\
\cmidrule(lr){1-1} \cmidrule(lr){2-4} \cmidrule(lr){5-7}
\textbf{Allen Relations} & \textbf{Explicit} & \textbf{Implicit} & \textbf{Temporal Answer} & \textbf{Explicit} & \textbf{Implicit} & \textbf{Temporal Answer} \\
\midrule
Empty & 42 & 3 & 13298 & 0 & 0 & 304 \\
After & 36929 & 527 & - & 877 & 10 & - \\ 
Before & 38058 & 262 & - & 937 & 3 & - \\ 
Contains & 816 & 2 & - & 23 & 0 & - \\ 
During & 17463 & 230 & - & 451 & 7 & - \\ 
Equals & 9856 & 102 & - & 223 & 2 & - \\ 
Finishes & 134 & 11 & - & 7 & 0 & - \\
FininshedBy & 11 & 0 & - & 0 & 0 & - \\
Meets & 98 & 10 & - & 1 & 0 & - \\
MetBy & 61 & 29 & - & 4 & 0 & - \\ 
Overlaps & 6559 & 40 & - & 149 & 1 & - \\
OverlappedBy & 49 & 1 & - & 0 & 0 & - \\
Starts & 513 & 7 & - & 15 & 0 & - \\
StartedBy & 17 & 7 & - & 0 & 0 & - \\
\midrule
Total & 110606 & 1223 & 13298 & 2687 & 23 & 304 \\   
\bottomrule
\end{tabular}
}
\caption{
    Augmented Temporal Nobel Prize extra annotations as byproduct of our augmentation pipeline.
}
\label{tab:tnp_allen_temporal_type_count}
\end{table*}
\begin{table*}[ht]
\centering
\scalebox{0.9}{
\begin{tabular}{lccccc}
\toprule
&
\multicolumn{3}{c}{\textbf{Train}} &
\multirow{2}{*}{\textbf{Dev}} &
\multirow{2}{*}{\textbf{Test}} \\
\cmidrule(lr){2-4}
Time Sensitive QA & Original & Original (Modified) & Our \\
\midrule
\# Unique Wikipedia documents & 3401 & 2774 & 2774 & 622 & 4926 \\
\# Passages & 9166 & 5143 & 8149 & 826 & 109182 \\
\# Positive queries & 26936 & 15051 & 44015 & 3162 & 5167 \\
\# Negative queries & 0 & 0 & 41926 & 0 & -\\
\# Qrel & - & - & - & - & 5247 \\
   
\bottomrule
\end{tabular}
}
\caption{
    Dataset statistics of the original, modified, and augmented TimeQA dataset.
}
\label{tab:timeqa_stats}
\end{table*}

\begin{table*}[ht]
\centering
\scalebox{0.9}{
\begin{tabular}{l|ccc}
\toprule

\textbf{TimeQA} & \multicolumn{3}{c}{\textbf{Train}} \\
\cmidrule(lr){1-1} \cmidrule(lr){2-4}
\textbf{Allen Relations} & \textbf{Explicit} & \textbf{Implicit} & \textbf{Temporal Answer} \\
\midrule
Empty & 19 & 116 & 8514 \\
After & 24224 & 320 & - \\
Before & 28188 & 235 & - \\
Contains & 1024 & 1 & - \\
During & 12081 & 134 & - \\
Equals &  6551 & 54 & - \\
Finishes & 121 & 2 & - \\
FininshedBy & 2 & 0 & - \\
Meets & 52 & 5 & - \\
MetBy & 23 & 12 & - \\
Overlaps & 3922 & 21 & - \\ 
OverlappedBy & 52 & 0 & - \\  
Starts & 250 & 7 & - \\
StartedBy & 10 & 1 & - \\ 
\midrule
Total & 76519 & 908 & 8514 \\   
\bottomrule
\end{tabular}
}
\caption{
    Augmented TimeQA extra annotations as byproduct of our augmentation pipeline.
}
\label{tab:timeqa_allen_temporal_type_count}
\end{table*}

\paragraph{Temporal Nobel Prize (TNP).}
We show the detailed statistics of the TNP dataset~\cite{wuCIKM24TimeSensitveRetrievalAugmented2024} and our augmented version in Table~\ref{tab:tnp_passage_query_count}.
Table~\ref{tab:tnp_allen_temporal_type_count} further details the statistics of temporal query types and Allen relations as a by-product of our Temporal Contrastive Learning augmentation pipeline in Section~\ref{sec:data}.

\paragraph{Time-sensitive QA (TimeQA).}
We present detailed statistics of the TimeQA dataset~\cite{chenNeurIPS21DatasetAnswering2021} and our augmented version in Table~\ref{tab:timeqa_stats}.
Table~\ref{tab:timeqa_allen_temporal_type_count} further reports the statistics of temporal query types and Allen relations as a byproduct of our Temporal Contrastive Learning augmentation pipeline described in Section~\ref{sec:data}.

TimeQA is a popular TQA benchmark constructed from Wikipedia, where high semantic similarity between queries and documents makes evaluating temporal encoding challenging~\cite{wuCIKM24TimeSensitveRetrievalAugmented2024}. 
To adapt TimeQA into a temporal information retrieval benchmark, we perform careful preprocessing to mitigate the issue of queries and passages being overly semantically aligned~\cite{wuCIKM24TimeSensitveRetrievalAugmented2024}.
We use the Wikipedia snapshot \texttt{enwiki-20211220-pages-articles}, which is the only publicly accessible snapshot with dates closest to those used in TimeQA.
We use \texttt{index\_enwiki-20220820} Wikipedia index database.
This snapshot is first processed following FlashRAG~\cite{jinWWW25FlashRAGModular2025}\footnote{\href{https://github.com/RUC-NLPIR/FlashRAG/blob/main/docs/original\_docs/process-wiki.md}{https://github.com/RUC-NLPIR/FlashRAG/blob/main/docs/original\_docs/process-wiki.md}},
and then filtered using SUTime to retain only pages that contain at least one temporal expression and include an answer from the TimeQA train, development, or test sets.

After splitting and generating our augmented dataset, we further filter the original training set to retain only passages whose Wikipedia IDs match those in the augmented dataset, resulting in the ``Original (Modified)'' column in Table~\ref{tab:timeqa_stats}. 
We use this training set as the baseline data for our augmentation pipeline. 
We do not modify the development set so that model selection is still evaluated under a semantically dominant retrieval setting.

To construct the test corpus, we combine all training, development, and test files from both the easy and hard splits.
We then use the index database to map the corresponding passages’ Wikipedia URLs to Wikipedia IDs.
All pages with non-null Wikipedia IDs from the selected snapshot are accumulated to form the test corpus.
To create the test qrel and corpus files, we retain only queries that are answerable.

Additionally, for each page, we use \texttt{chonkie}~\cite{chonkie2025} to further chunk the filtered corpus into smaller paragraphs using the \texttt{o200k\_base} tokenizer with a chunk size of 128.
Finally, for each test query, we prompt \texttt{Qwen/Qwen3-4B-Instruct-2507} against all chunked paragraphs from the corresponding Wikipedia pages, assisted by SUTime filtering and manual inspection.
This process ultimately results in the test set summarized in Table~\ref{tab:timeqa_stats}.
The prompts used for qrel construction are provided in Table~\ref{tab:timeqa_qrel_prompt}.

\begin{table*}[t]
\footnotesize
\scalebox{0.95}{
\begin{tabularx}{\textwidth}{X}
\toprule
\textbf{Prompt template for temporal annotation}

You are a temporal annotation expert for information retrieval. We provide you with a query that may contain one or many temporal expressions and the corresponding  SUTIME's output for your reference. We also provide you with previously annotated examples of positive and negative samples of the query; use them as references if they are semantically relevant to the query. Your job is to generate high-quality positive passages and negative passages for temporal contrastive learning.
Your final output are annotated JSONs (no indentation) that strictly follow the provided JSON schema.

Your temporal annotations must be more precise and reliable than SUTIME or HeidelTime, with top-tier accuracy. You must follow the definitions and instructions below.

* Allen relations and descriptions: \textbf{<Allen relations and descriptions>}

* Instructions:

1. You must output one or multiple valid JSONs, delimited by a newline, strictly following this pydantic schema: \textbf{<Pydantic JSON Schema>}

2. "query\_id": must be assigned with the provided "query\_id".

3. “query":
- A query is a provided sentence that may contain one or many temporal expressions.
- Please note that SUTIME only provides explicit temporal expressions for the query and they are not perfect; therefore, you must double-check them and further detect additional explicit and implicit temporal expressions such as events.
- You must extract all events that are anchored to specific datetime (e.g., “2000 FA Cup Final”, “the 2007 election”, etc.).
- The "temporal" field should be extracted as written from the query's text and they must be concise, such as: "in April, 1906", "2 July 2010", "2004 general election", etc.

4. "positive\_passages":
- Natural, QA-style questions with diverse phrasing that seek information from the query. Each question must contain exactly one extractable temporal expression, logically align with one of the query's temporal expressions, yet be diverse in Allen relations.
- Based on the query's temporal expressions, you must generate "positive\_passages" that cover all "TemporalQueryType", including 'Explicit', 'Implicit', and 'TemporalAnswer', when possible:
    - You MUST prioritise generating questions with explicit temporal constraints, like "in 2010", "from 2010 to 2015", etc. They must logically align with the query’s temporal expression(s), such as being equals, overlapping with, or being contained within the query’s temporal expression(s). If the query has multiple temporal expressions, the generated questions must logically align with at least one of them.
    - You should also prioritise generating questions with implicit temporal constraints, such as "after EVENT", "before EVENT". These must logically align with the query’s temporal expression(s).
    - For both explicit and implicit "TemporalQueryType" questions, you must not use phrases like what/which date/day/month/year/time or when etc., that inquire about time. You must not confuse this with TemporalAnswer questions.
    - You can also generate "TemporalAnswer" (asking for datetime/duration/time-range of an EVENT, e.g., "When did EVENT happen?", "What time did he arrive?"). For this type of question, you must not add any temporal expression and you must set: "TemporalQueryType": "TemporalAnswer", "allen\_relation": "Empty", and "temporal": []. Use sparingly (less than 2 per query, only when explicit/implicit cannot be formed).
- "temporal" field: must extract all exact text spans of the temporal expressions that exist in the generated question (not normalized or paraphrased). Regarding explicit and implicit "TemporalQueryType", they must be concise temporal expressions as written, such as "after July 2010", "from 2012 to 2014", that are not just normalized dates. For instance, instead of "In 1906 he moved with his family to a farm", prefer the concise "In 1906" and keep prepositions or context words that anchor the time, e.g., 'from', 'in', 'after', etc. Regarding "TemporalAnswer" passages, the "temporal" field must be an empty list.
- Allen relation: consider Passage = Event A and Query = Event B. Must be correct and align with the provided definition.
- “docid”: must remain the same as provided.
- Quantity: For each split/rewritten query, you must generate a list of 5 high-quality positive passages, prioritizing quality over quantity.

5. "negative\_passages": follow the same format as "positive\_passages", but represent contexts that are temporally mismatched or semantically irrelevant to the query. Based on the query's temporal expressions, you must generate "negative\_passages" that cover all "TemporalQueryType", including "Explicit", "Implicit", and "TemporalAnswer", when possible. They must be hard, temporally-confused, yet diverse in Allen relations questions that cover all following cases:
- Case 1: Questions with temporal expressions that mismatch with the query’s temporal expression(s): 
    - If the query specifies a span (e.g., 2005–2007), use an interval that falls completely outside it (e.g., 2008, before 2005, after 2007).
    - Adjacent years or ranges are valid negatives (e.g., query = 2010, negative = 2009).
    - Use shifted but non-overlapping intervals (e.g., query = 2010, negative = 2012–2014).
    - Include misleading implicit cues (e.g., “shortly after 2011” vs. query “in 2010”).
    - You may reuse the same passage from "positive\_passages" but replace its temporal expressions with mismatched ones.
- Case 2: Questions with same temporal expression but irrelevant event/entity
    - Questions with overlapping or identical temporal expressions but targeting a different subject.
    - Example: query = “Ronaldo’s career in 2010” vs. negative = “Messi’s career in 2010”.
    - This ensures negatives are temporally aligned but semantically irrelevant.
- Case 3: "TemporalAnswer"-type questions that are either:
    - Irrelevant to the query.
    - Or ask for non-existent temporal information in the query context.
    - Use sparingly (less than 2 per query, only when explicit/implicit cannot be formed).
- Allen relation: consider Passage = Event A and Query = Event B. Must be correct and align with the provided definition.
- Quantity: For each split/rewritten query, you must generate a list of 5 high-quality negative passages, prioritizing quality over quantity.  
- Do not create trivial negatives (e.g., completely unrelated random text).
- Ensure no accidental overlap with valid facts in the document.

6. "temporal\_query\_type": Must be "Explicit", "Implicit", or "TemporalAnswer". If the temporal contains a clear date/month/year, it is "Explicit", NOT "Implicit". If the passage asks for which date/month/year of an event, it is "TemporalAnswer".
   
7. Final output: Only output valid JSON(s). Do not explain, add comments, or include extra text, since your output will be parsed automatically. 

\#\#\# Demonstration 1 \\
<Demo 1> \\
\#\#\# End of Demonstration 1 \\

\#\#\# Demonstration 2 \\ 
<Demo 2> \\ 
\#\#\# End of Demonstration 2
\\
\bottomrule
\end{tabularx}
}
\caption{
Detailed prompts for augmenting the positive/negative queries. 
The Allen relations and Pydantic JSON schemas are defined in Table~\ref{tab:allen_relation} and Table~\ref{tab:json_schema}, respectively.}
\label{tab:prompt_template1}
\end{table*}

\begin{table*}[t]
\begin{tabularx}{\textwidth}{X}

\toprule
\textbf{Allen relations' descriptions}

- Before: Did ‘Event A’ occur before ‘Event B’ without any overlap between the two events?

- After: Did ‘Event A’ occur after ‘Event B’ without any overlap between the two events?

- Meets: Did ‘Event A’ end in the same time as ‘Event B’ began? Answer True or False.

- MetBy: Did ‘Event B’ end in the same time as ‘Event A’ began? Answer True or False.

- Overlaps: Did ‘Event A’ begin before ‘Event B’ and end before ‘Event B’ ended, with some overlap between the two events?

- OverlappedBy: Did ‘Event B’ begin before ‘Event A’ and end before ‘Event A’ ended, with some overlap between the two events?

- Starts: ‘Event A’ begin in the same time as ‘Event B’, but end before ‘Event B’ ended?

- StartedBy: Did ‘Event B’ begin in the same time as ‘Event A’, but end before ‘Event A’ ended?

- During: Did ‘Event A’ begin after ‘Event B’ began and end before ‘Event B’ ended, being entirely contained within ‘Event B’?

- Contains: Did ‘Event A’ begin before ‘Event B’ began and end after ‘Event B’ ended, entirely containing ‘Event B’?

- Finishes: Did ‘Event A’ begin after ‘Event B’ began and end in the same time as ‘Event B’?

- FinishedBy: Did ‘Event B’ begin after ‘Event A’ began and end in the same time as ‘Event A’?

- Equals: Did ‘Event A’ begin in the same time as ‘Event B’ and end in the same time as ‘Event B’?

- Empty: a special case for TemporalAnswer where there is no temporal expression.
\\
\bottomrule
\end{tabularx}
\caption{Detailed prompts for Allen relations descriptions. Except for ``Empty'', the descriptions are provided by~\citet{islakoglu-kalo-2025-chronosense}.}
\label{tab:allen_relation}
\end{table*}
\begin{table*}[t]
\begin{tabularx}{\textwidth}{X}

\toprule
\textbf{Pydantic JSON schema}

{`\$defs': {`AllenRelation': {`enum': [`Before', `After', `Meets', `MetBy', `Overlaps', `OverlappedBy', `Starts', `StartedBy', `During', `Contains', `Finishes', `FinishedBy', `Equals', `Empty'], `title': `AllenRelation', `type': `string'}, `Passage': {`properties': {`docid': {`title': `Docid', `type': `integer'}, `text': {`title': `Text', `type': `string'}, `temporal': {`items': {`type': `string'}, `title': `Temporal', `type': `array'}, `temporal\_query\_type': {`\$ref': `\#/\$defs/TemporalQueryType'}, `allen\_relation': {`\$ref': `\#/\$defs/AllenRelation'}}, `required': [`docid', `text', `temporal', `temporal\_query\_type', `allen\_relation'], `title': `Passage', `type': `object'}, `TemporalQueryType': {`enum': [`Explicit', `Implicit', `TemporalAnswer'], `title': `TemporalQueryType', `type': `string'}}, `properties': {`query\_id': {`title': `Query Id', `type': `integer'}, `query': {`title': `Query', `type': `string'}, `temporal': {`items': {`type': `string'}, `title': `Temporal', `type': `array'}, `positive\_passages': {`items': {`\$ref': `\#/\$defs/Passage'}, `title': `Positive Passages', `type': `array'}, `negative\_passages': {`items': {`\$ref': `\#/\$defs/Passage'}, `title': `Negative Passages', `type': `array'}}, `required': [`query\_id', `query', `temporal', `positive\_passages', `negative\_passages'], `title': `TemporalAnnotation', `type': `object'}.
\\
\bottomrule
\end{tabularx}
\caption{Detailed prompts for Pydantic JSON schema}
\label{tab:json_schema}
\end{table*}
\begin{table*}[t]
\footnotesize
\scalebox{0.95}{
\begin{tabularx}{\textwidth}{X}
\toprule
\textbf{Prompt template for temporal annotation}\\
Input format
(index, query, list of answer\_terms, passage)
\\
Task\\
Decide whether the passage is relevant to the query for temporal information retrieval (qrel construction).
A passage is relevant if and only if ALL of the following conditions are satisfied:
\\
1. Temporal correctness (mandatory): The passage must clearly satisfy the temporal constraint(s) explicitly or implicitly expressed in the query (e.g., in, before, after, between, from … to …).
- Explicit dates or time spans in the passage are acceptable.\\
- Implicit temporal coverage is acceptable only if it unambiguously covers the queried time period (e.g., “served from 1989–1993” satisfies “in 1991”).\\
- Vague, approximate, or underspecified temporal expressions (e.g., “around that time”, “in the early days”, “in the 1900s”) do not satisfy temporal constraints.\\
- If the temporal alignment between the query and passage is unclear, the passage must be judged not relevant.\\
2. Answer-term validity within time: The passage must contain all answer\_terms.
- The temporal information must be directly associated with the answer\_terms, meaning the answer\_terms refer to the same entity, event, or state that is valid within the specified time frame. A passage is relevant if and only if ALL of the following conditions are satisfied:
\\
- Merely mentioning the answer\_terms elsewhere in the passage without temporal grounding is insufficient.\\
3. No external inference: Judge relevance only based on the information explicitly present in the passage.\\
- Do not rely on external knowledge, common facts, or assumptions.\\
- Do not infer missing dates or temporal relations.\\
4. Binary decision rule: If any of the above conditions is not satisfied, the passage is not relevant.\\

Output format \\
- Output the integer index only if the passage is relevant.\\
- Otherwise, output -1.\\
- Output only the index or -1.\\
- Do not include explanations, reasoning steps, or additional text.\\
- Do not guess or assign partial relevance.\\

\#\#\# Demonstration 1\\
Input:\\
(Index: 123, Question: Who was the head coach of the team 1. FC Köln from Nov 2019 to Nov 2020?, Answers: ['Markus Gisdol'], Passage: "'Section: Decline and changes (2018–). FC Saarbrücken, the club decided to terminate Beierlorzer's contract on 9 November 2019. Sporting director Armin Veh, who weeks earlier had announced that he would not extend his contract with the club, was also dismissed from his position. On 18 November, former HSV manager Markus Gisdol was appointed to the club's head coaching position, while Horst Heldt was made sporting director.  Both signed contracts until 2021. After avoiding relegation at the end of the season, Gisdol's contract was extended until 2023. )"\\
Output: 123\\
\#\#\# End of Demonstration 1\\

\#\#\# Demonstration 2\\
Input:\\
(Index: 2, Question: Who was the head coach of the team 1. FC Köln from Nov 2019 to Nov 2020?, Answers: ['Markus Gisdol'], Passage: "Section: Decline and changes (2018–).   The club found itself in a renewed relegation during the 2020–21 season. On 11 April 2021, after losing to relegation rival Mainz 05, Gisdol was dismissed from his position as head coach. The next day, it was announced that Friedhelm Funkel would take over head coaching duties until the end of the season. On 11 May, it was reported that SC Paderborn manager Steffen Baumgart would succeed Funkel as head coach at the beginning of the 2021–22 season. )"\\
Output: -1\\
\#\#\# End of Demonstration 2\\

\#\#\# Demonstration 3\\
Input:\\
(Index: 3, Question: Who was the head coach of 1. FC Köln from Nov 2019 to Nov 2020?, Answers: ['Markus Gisdol'], Passage: "'Gisdol was appointed head coach in November 2018 and left in October 2019 before the season ended.'")\\
Output: -1\\
\#\#\# End of Demonstration 3 \\
\bottomrule
\end{tabularx}
}
\caption{
Detailed prompts for generating the TimeQA's test qrel file.}
\label{tab:timeqa_qrel_prompt}
\end{table*}

\end{document}